\documentclass[10pt,journal,compsoc]{IEEEtran}
\ifCLASSOPTIONcompsoc
  \usepackage[nocompress]{cite}
\else
  \usepackage{cite}
\fi
\ifCLASSINFOpdf
\else
\fi
\usepackage{booktabs} 

\usepackage[english]{babel}
\usepackage{moresize}
\usepackage{amsmath}

\usepackage{algorithmic} 
\usepackage{balance}
\usepackage{multirow}
\usepackage{comment}
\usepackage{paralist}

\usepackage[english]{babel}
\usepackage[latin1]{inputenc}
\usepackage{mathrsfs}
\usepackage{graphicx}
\usepackage{amssymb}
\usepackage{url}
\usepackage{mathrsfs}
\usepackage{subfigure}
\usepackage{amsmath}
\usepackage{enumitem}
\usepackage[linesnumbered,algoruled,boxed,lined]{algorithm2e}
\usepackage{adjustbox}
\usepackage[justification=centering]{caption}

\usepackage{pgfplots}
\usetikzlibrary{pgfplots.dateplot}

\newcommand{\hide}[1]{} 

\newcommand{\etal}{\textit{et al}.}

\newcommand{\ie}{\textit{i}.\textit{e}.}
\newcommand{\eg}{\textit{e}.\textit{g}.}

\def\model{TGT}

\begin{document}
%


\title{Multi-Behavior Sequential Recommendation with Temporal Graph Transformer}


\author{Lianghao Xia,
	Chao Huang$^*$,
	Yong Xu,
	and Jian Pei, \emph{Fellow, IEEE}
	\IEEEcompsocitemizethanks{\IEEEcompsocthanksitem L. Xia, C. Huang are with the Department of Computer Science \& Musketeers Foundation Institute of Data Science, at the University of Hong Kong, Hong Kong, China. Email: aka\_xia@foxmail.com, chaohuang75@gmail.com.
	\IEEEcompsocthanksitem Y. Xu is with the School of Computer Science and Technology, South China University of Technology, Guangzhou, China. Email: yxu@scut.edu.cn.
	\IEEEcompsocthanksitem J. Pei is with School of Computing Science, Simon Fraser University. E-mail: jpei@cs.sfu.ca.
	\IEEEcompsocthanksitem Chao Huang is the corresponding author.}
}

\IEEEtitleabstractindextext{%

\begin{abstract}
Modeling time-evolving preferences of users with their sequential item interactions, has attracted increasing attention in many online applications. Hence, sequential recommender systems have been developed to learn the dynamic user interests from the historical interactions for suggesting items. However, the interaction pattern encoding functions in most existing sequential recommender systems have focused on single type of user-item interactions. In many real-life online platforms, user-item interactive behaviors are often multi-typed (\eg, click, add-to-favorite, purchase) with complex cross-type behavior inter-dependencies. Learning from informative representations of users and items based on their multi-typed interaction data, is of great importance to accurately characterize the time-evolving user preference. In this work, we tackle the dynamic user-item relation learning with the awareness of multi-behavior interactive patterns. Towards this end, we propose a new \underline{T}emporal \underline{G}raph \underline{T}ransformer (\model) recommendation framework to jointly capture dynamic short-term and long-range user-item interactive patterns, by exploring the evolving correlations across different types of behaviors. The new \model\ method endows the sequential recommendation architecture to distill dedicated knowledge for type-specific behavior relational context and the implicit behavior dependencies. Experiments on the real-world datasets indicate that our method \model\ consistently outperforms various state-of-the-art recommendation methods. Our model implementation codes are available at https://github.com/akaxlh/TGT.
\end{abstract}


}

\maketitle

\section{Introduction}
\label{sec:intro}

Personalized recommender systems have become increasingly important to alleviate information overload and satisfy user diverse interests in many online platforms (\eg, E-commerce sites, online movie and music sites)~\cite{zhao2020catn,2019online,hansen2020contextual}. While many collaborative filtering methods have been proposed to model user-item interactions~\cite{he2017neural,xue2019deep}, they focus on the static settings and ignore the dynamic nature of user's time-evolving preferences~\cite{liu2018stamp}. Hence, among various recommendation scenarios, sequential recommender systems are introduced to model dynamic preference of users based on their sequential interaction behaviors~\cite{sun2019bert4rec,chorus2020,ma2020aaai}. The key idea of sequential recommendation models is to understand the evolution of user's preference by capturing temporal dependency of user-item interactions, based on the past observed behaviors.

With the advent of deep learning techniques in recent years, a substantial number of approaches have been proposed to solve the sequential recommendation problem, via utilizing different neural network techniques. For example, recurrent neural network-based models propose to encode the sequential information from the interacted item sequence of each user~\cite{hidasi2015session,quadrana2017personalizing}. In addition, convolutional neural networks and attention mechanism also serve as effective solutions for modeling sequential patterns of item-item transitions. \eg, Caser~\cite{tang2018personalized} designs convolution-based kernel functions to aggregate information from neighboring time slots. NARM~\cite{li2017neural} and SASRec~\cite{kang2018self} develop attentive relation encoder to capture users' general interests with long-term temporal dependency. Inspired by the effectiveness of graph neural networks, several GNN-based methods (\eg, HyRec~\cite{wang2020next}, MA-GNN~\cite{ma2020aaai} and MTD~\cite{huang2021graph}) exploit the user-item graph structure to guide the embedding learning with the incorporation of temporal context of user interactions.

\begin{figure}[t]
    \centering
    \includegraphics[width=0.48\textwidth]{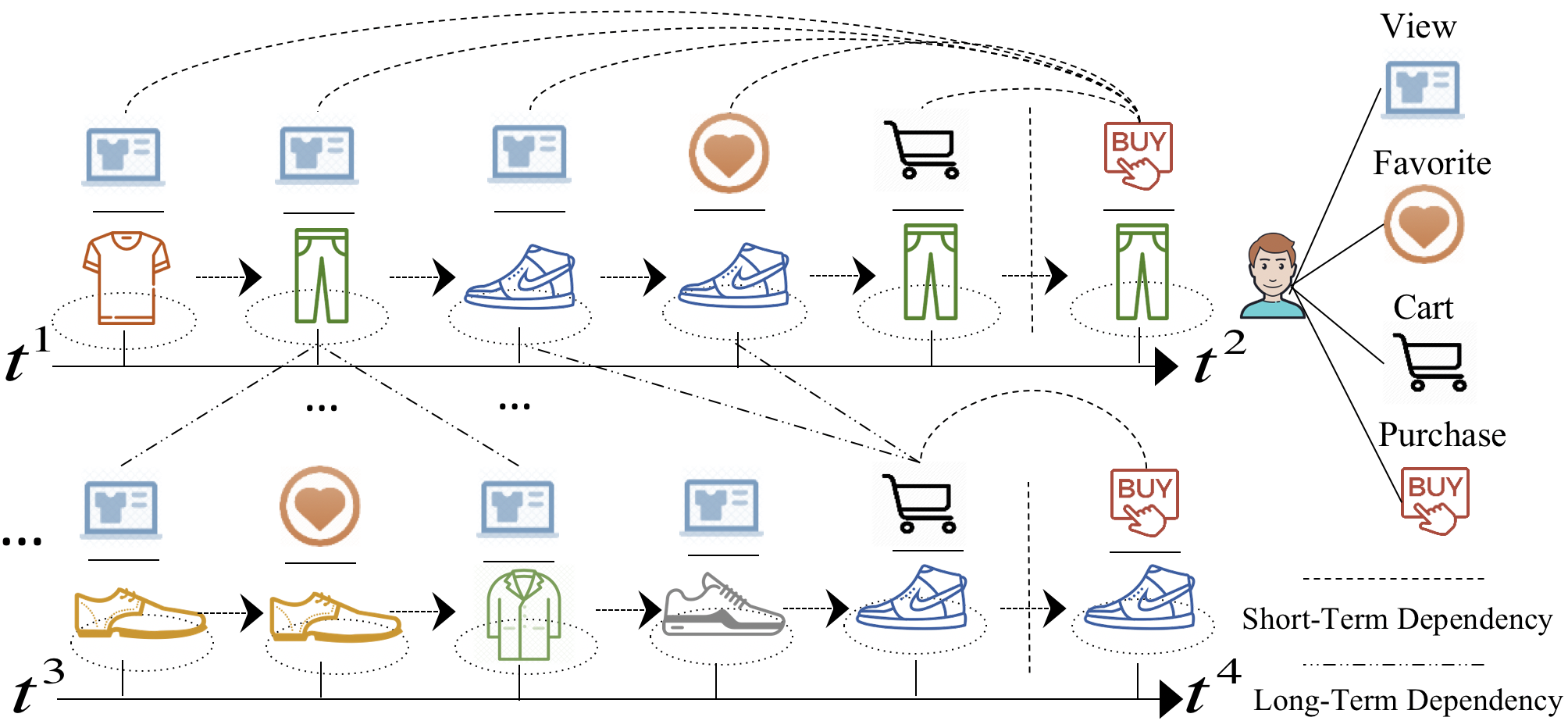}
    \caption{Illustration example of sequential recommendation with multi-behavior dynamics. Best viewed in color.}
    \label{fig:intro}
\end{figure}

While the aforementioned methods have achieved compelling results, we argue that there is a common deficiency in them: most of current frameworks are designed for single behavior type. In practice, the intention of user-item interactions can change over time depending on different contexts~\cite{tanjim2020,jin2020multi}. Let us consider an example from online retailing platforms: users could view products or tag them as favourites if they like, or make final purchase if the products meet their needs~\cite{gao2019neural} (as illustrated in Figure~\ref{fig:intro}). Hence, user-item interactions are often exhibited with time-dependent and behavior diversity in nature. To recommend the future purchase for a user, it is important and beneficial to explore not only what he/she has purchased before, but also what products this user has viewed previously or tagged them as his/her favorite items~\cite{xia2021knowledge}. In this work, we propose to capture the time-evolving user preference with the modeling of behavior heterogeneity and the underlying dependency in a dynamic environment, so as to improve the performance of sequential recommender systems.

However, exploring relation dependencies behind multi-behavior user-item interactions is intrinsically difficult, especially when the evolution of user interest is incorporated into the recommendation framework. Specifically, we face two key challenges: i) \textbf{Cross-type Behavior Inter-dependency}. Different types of user behaviors (\eg, click, tag as favorite, add-to-cart) may offer complementary signals for predicting the target behavior (\eg, purchase)~\cite{gao2019neural,jin2020multi}. If feature representations are learned from different behavior types separately and then loosely concatenate embeddings together, it can hardly capture the complex inter-dependency across various types of interaction behaviors. ii) \textbf{Temporal Multi-behavior Pattern Fusion}. Distinct to
stationary inter-dependent relations, it is challenging to be responsive for the evolution of multi-behavior semantics and the underlying cross-type behavior dependency. In practical applications, users often interact with items in a variety of behaviors which are inherently correlated, due to his/her specialty at different timestamps. Therefore, to built effective multi-behavior sequential recommendation model, it requires the careful design to jointly distill the behavior heterogeneity and underlying type-dependent patterns.\\\vspace{-0.1in}

In light of these challenges, we propose a multi-behavior sequential recommendation model with \underline{T}emporal \underline{G}raph \underline{T}ransformer (short for \model). In particular, to capture the short-term multi-typed interaction patterns of users, we develop a behavior-aware transformer network to inject the behavior heterogeneous signals into the sequential modeling of item transitions. In this regard, we preserve the expressive multi-behavior characteristics and the changes in interaction semantics. To exploit the long-term multi-behavior dependencies, we propose a temporal graph neural network to infer the latent user representations from their diversified activities on items with arbitrary durations. Instead of parameterizing each type of user-item interactions into separate embedding spaces, our designed multi-channel augmented message passing paradigm allow behavior of different types to maintain not only their specific time-aware semantics, but also the behavior type-specific dependent representations with long-range dynamics. Furthermore, in \model, we recursively refine the global-level representations over the time-aware user-item interaction graph to capture the dynamic cross-sequence correlations among different users. This has been largely overlooked by most of existing sequential recommender systems~\cite{sun2019bert4rec,wang2020next,ma2020aaai} for simplifying the model design.

We summarize the contributions of this work as follows:

\begin{itemize}[leftmargin=*]
    \item This work tackles the multi-behavior sequential recommendation with the exploration of multi-behavior characteristics of users from both the short-term and long-term perspectives. We show that modeling the dynamic cross-type behavior inter-dependency is essential for improving the recommendation quality.\\\vspace{-0.1in}
    \item We propose a general sequential recommendation model \model\ to maintain the dedicated representations for different types of user-item interactions over time. In \model, we adapt the graph neural network along the temporal dimension to capture the dynamism of  multi-behavior interaction patterns.\\\vspace{-0.1in}
    \item We conduct empirical studies on two real-world datasets to demonstrate the superiority of our \model\ framework when competing with various 18 recommendation baselines. Our evaluation also justifies the effectiveness of components in our \model\ model, as well as the interpretability of the learned multi-behavior dependent representations.
\end{itemize}

The remainder of this paper is organized as follows. Section~\ref{sec:relate} discusses the related work. In Section~\ref{sec:solution}, we introduce the details of our proposed framework \model. We report our experimental results that demonstrate the effectiveness of the proposed model in Section~\ref{sec:eval}. Finally, Section~\ref{sec:conclusion} concludes our study and discusses the future work.

\section{Related Work}
\label{sec:relate}

In this section, we summarize the relevant research work from the following research lines: \ie, i) sequential recommendation, ii) graph-based recommender systems, iii) recommendation with heterogeneous relational context, iv) multi-behavior/intent recommendation.

\subsection{Sequential Recommender Systems}
The sequential recommendation is usually formulated as sequence prediction problem with the modeling of item transitional regularities~\cite{wang2019sequential,huang2018improving,zhou2020s3}. Some earlier studies are on the basis of Markov Chain to model correlations between different items, such as factorizing personalized Markov Chains (FPMC)~\cite{rendle2010factorizing} and fusing item similarity with Markov Chains (Fossil)~\cite{fusingicdm}. To tackle the challenge of sequential behavior learning, many neural network-based methods have been proposed to capture complex sequential information~\cite{fang2020deep}. In particular, recurrent neural network is utilized as an intuitive way to model temporal dependencies~\cite{hidasi2015session}. Attention mechanisms allows the model to identify specific parts of input item sequences and preserve long-term dependencies, \eg, NARM~\cite{li2017neural} and SASRec~\cite{kang2018self}. For example, SASRec relies on the self-attention mechanism to identify item relevance from user past behaviors~\cite{kang2018self}. In addition, SR-GNN~\cite{wu2019session}, MTD~\cite{huang2021graph} and GCGNN~\cite{wang2020global} convert the sequential item transitions into graphs and design graph neural networks for item representations. However, these models only consider singular type of sequential behaviors. Given users' interacted item sequence, how to effectively exploit dynamic multi-behavior knowledge to improve recommendation accuracy, remains a significant challenge.

\subsection{Graph Neural Network for Recommendation}
Motivated by the strength of graph neural network, there exist many recently proposed recommendation methods modeling user-item interaction graph via information propagation between connected user/item nodes~\cite{zhu2020bilinear,wang2019neural,fan2019metapath,zhang2022dynamic,xia2022hypergraph}. For example, Wang~\etal~\cite{wang2019neural} and Ying~\etal~\cite{ying2018graph} introduces the graph convolutional network in modeling collaborative effects between users and items. The embedding propagation is conducted over the constructed user-item interaction graph. Later on, He~\etal~\cite{he2020lightgcn} propose to improve training efficiency by removing the feature transformation and nonlinear activation operations with the light graph convolutional framework. With the self-supervised learning techniques, SGL~\cite{wu2021self} performs data augmentation on user-item interaction graphs to generate additional supervision signals. HCCF~\cite{xia2022hypergraph} enhances the graph-based collaborative filtering with hypergraph-enhanced contrastive learning. In this work, our \model\ framework is built upon the dynamic graph neural architecture to capture the high-order connectivity, with the exploration of evolving multi-behavior preference.


\subsection{Recommendation with Heterogeneous Relations}
Another relevant research line is to consider different types of relationships in recommender systems~\cite{2021recent,jin2020efficient}. First, with the emergence of online social networks, social-aware recommendation models have been proposed to jointly consider user-user and user-item relations for better user embedding generation~\cite{fan2019graph,wu2019neural,huang2021knowledge}. Second, recent studies attempt to incorporate knowledge graph relational information as side features to model item correlations~\cite{ma2019jointly,chen2020jointly,cao2019unifying}. Third, to further alleviate data sparsity issue, heterogeneous graph learning approaches aims to transfer knowledge from both user and item domains (\eg, social links, item semantic relatedness), with the goal of augmenting the data of user-item pairs~\cite{shi2018heterogeneous,liu2020heterogeneous}. Different from those methods which rely on the the side information to construct meta-relations between users and items, this work focuses on exploiting the dynamic characteristics of multi-behavioral interaction data and validates its positive effects in sequential recommendation.

\subsection{Multi-Behavior/Intent Recommender Systems}
There exist some recommendation methods which aim to learn the collective knowledge from different types of user-item interactions (\eg, clicks, purchases)~\cite{gao2019neural,jin2020multi,xia2021graph,wei2022contrastive}. For example, Gao~\etal~\cite{gao2019neural} define the cascading relationship to account for multiple types of user behaviors. AIR~\cite{chen2019air} is an attention-based recurrent neural model to capture the intention-aware sequential patterns. In Intention2Basket~\cite{wang2020intention2basket}, heterogeneous user intentions are considered to encode the dynamic user preference for next-basket planning. Additionally, RIB~\cite{zhou2018micro} proposes to learn the sequential patterns from the perspective of micro behaviors, so as to improve the recommendation performance. In a session-based recommendation scenario, Meng~\etal~\cite{meng2020incorporating} introduces a new recommender system MKM-SR to integrate the item knowledge and different types of user behaviors for capturing item transitional patterns under a multi-task learning framework. MATN~\cite{xia2020multiplex} is an attention-based recommendation method which fuses behavior-aware patterns to generate weighed summarized representations. Additionally, based on graph neural networks, Jin~\etal~\cite{jin2020multi} uses graph convolutional network to model behavior-aware collaborative signals. CML~\cite{wei2022contrastive} designs a self-supervised learning method for multi-behavior recommendation. However, those models are designed for static recommendation scenarios, and can hardly be adaptive for modeling user-item interactions which are inherently dynamic. To fill this gap, we design a dedicated temporal graph-contextualized transformer network for modeling multi-behavior sequential interactions. Additionally, disentangled representation learning has been leveraged to separate embedding into distinct factors~\cite{ma2019learning,wang2020disenhan,zheng2021disentangling}. Nevertheless, most of existing disentangled representation works force the embedding separation with the assumption of implicit factor independence. Different from them, this work exploits the dynamic dependency across different types of behavioral interaction patterns in an explicit manner.

\section{Methodology}
\label{sec:solution}
\begin{figure*}
    \centering
    \includegraphics[width=\textwidth]{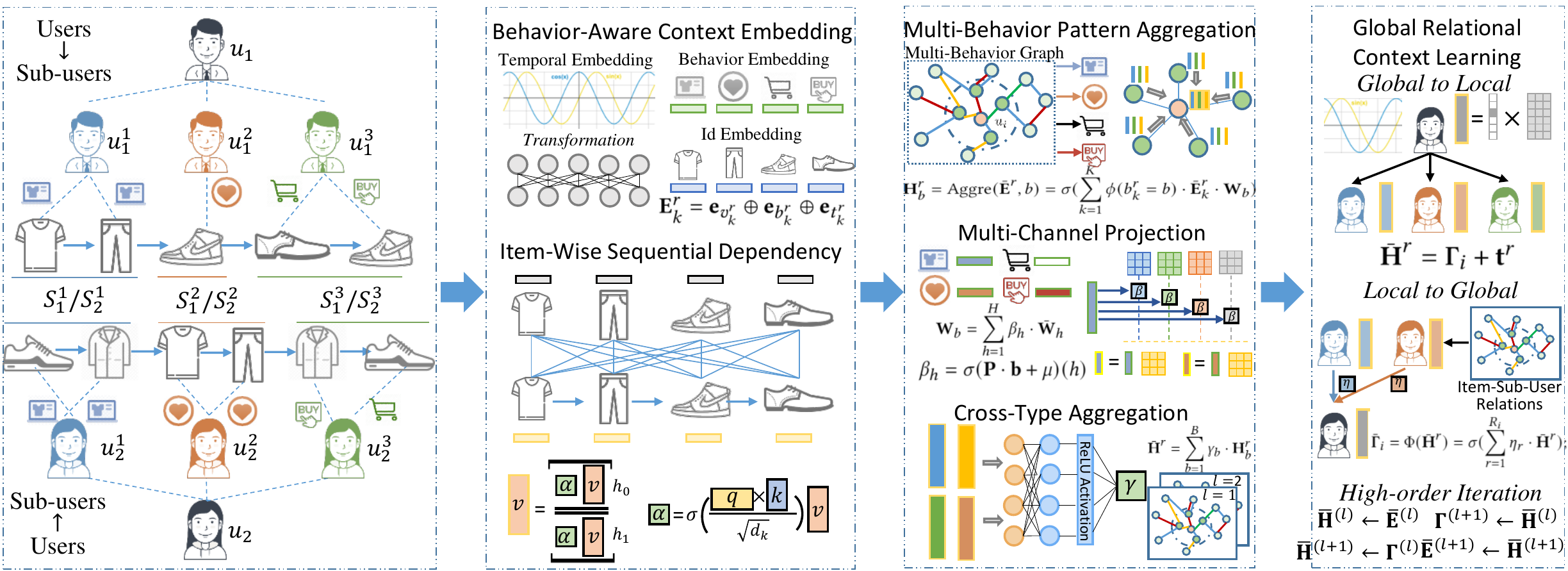}
    \vspace{-0.1in}
    \caption{Model flow of the proposed \model. (a) Hierarchical graph structure between user and item with respect to both short-term and long-term multi-behavior relations. (b) Dynamic individual interest modeling with behavior-aware sequential context. (c) Multi-behavior pattern aggregation which jointly preserves behavior semantics and cross-type behavior dependency. (d) Global relational context injection with the high-order embedding propagation paradigm.}
    \label{fig:framework}
    \vspace{-0.1in}
\end{figure*}

\begin{table}[t]
    \caption{Descriptions of Key Notations}
    \label{tab:data}
    \centering
    \footnotesize
	\setlength{\tabcolsep}{0.6mm}
    \begin{tabular}{c | c}
        \hline
        Notations & Description \\
        \hline
        $u_i\in \mathcal{U}$ & User set \\
        $v_j \in \mathcal{V}$  & Item set \\
        $b\in \mathcal{B}$ & Behavior types (indexed by $b$) \\
        $S_i$ & Multi-behavior interaction sequence \\
        $(v_{k}^r, t_{k}^r, b_{k}^r)$ & Interaction instance in the $r$-th sub-sequence \\
        $\textbf{e}_{v_k^r}$, $\textbf{e}_{b_k^r}$, $\textbf{e}_{t_k^r}$ & Embedding of item, behavior type and time \\
        $\bar{\textbf{E}}_k^r \in\mathbb{R}^d$ & Embedding of the $k$-th interacted item  \\
        $\alpha_{k,k'}$ & item relevance weight between $v_k^r$ and $v_{k'}^r$ \\
        $G_r$ & User sub-sequence interaction graph \\
        $G$ & Graph over users and their time-aware interactions \\ 
        $\bar{\textbf{H}}^r \in\mathbb{R}^d$ & Time-aware multi-behavior context representation \\
        $\bar{\mathbf{\Gamma}}_i \in\mathbb{R}^d$ & Cross-time global user embedding \\
        \hline
    \end{tabular}
    \label{tab:notations}
\end{table}

In this section, we first introduce the problem formulation and then elaborate our proposed \model\ framework. The model flow of our method is illustrated in Figure~\ref{fig:framework}. We also summarize key notations of our methodology in Table~\ref{tab:notations}.

\subsection{Task Formulation}
In our recommendation scenario, we let $\mathcal{U}=\{u_1,...,u_i,...,u_I\}$ denote the set of users and $\mathcal{V}=\{v_1,...,v_j,...,v_J\}$ denote the set of items. We further define $\mathcal{B}$ (indexed by $b$) to represent the types of different interaction behaviors, such as browsing, adding to cart, tagging as favorite and purchasing. We differentiate users' interactions by associating them with the corresponding behavior types ($b\in \mathcal{B}$). \\\vspace{-0.1in}

Definition 1. \textbf{Multi-behavior Interaction Sequence $S_i$}. Given a user $u_i\in \mathcal{U}$, his/her behavior-aware interaction sequence $S_i$ is consisted of individual triple $(v_{k,i}, b, t)$, representing the $k$-th interacted item (ordered by time) under the $b$-th behavior type at time $t$, \ie, $S_i=[(v_{1,i}, t_{1,i}, b_{1,i}),..., (v_{k,i}, t_{k,i}, b_{k,i}), ...,(v_{K,i}, t_{K,i}, b_{K,i})]$, where $K$ (indexed by $k$) is the sequence length $K=|S_i|$. \\\vspace{-0.1in}

In our multi-behavior sequential recommendation problem, we define the behavior type we aim to forecast as \emph{target behavior} (\eg, purchase in online retailing platforms). Other types of user behaviors (\eg, page view, tag-as-favorite) are considered as \emph{context behavior} for characterizing the preference of users over the target behavior. This work aims to predict the future item that user $u_i$ will be interested after the observed interaction sequence $S_i$, with the exploration of multi-behavior sequential patterns and interaction relation heterogeneity. Formally, our studied problem can be formalized as: \textbf{Input}: the past multi-behavior interacted item sequence of each user $S_i$ ($u_i\in U$); \textbf{Output}: a learning function which accurately predicts the interacted item of each user after time $t_{K,i}$.

\subsection{Framework Overview}
Our proposed \model\ is an integrative multi-behavior sequential recommendation framework which consists of several key learning phases (as shown in Figure~\ref{fig:framework}): 

\begin{itemize}[leftmargin=*]

\item To capture the behavior-aware sequential dependency across user-item interactions, we design an integrative relational learning framework with a behavior-aware context embedding module and an item-wise sequential dependency encoder. This component is able to model the temporal interaction patterns of users with the preservation of multi-behavior contextual signals.\\\vspace{-0.12in}

\item The component of multi-behavior pattern aggregation is proposed to capture short-term user preference over behavior-aware sub-interactions. In this module, the time-aware user-item interactive patterns are encoded by the dual-stage behavior-aware message passing paradigm for embedding propagation and refinement.\\\vspace{-0.12in}

\item Finally, a global context learning component is introduced to endow our \model\ with the ability of incorporating the long-term multi-behavior patterns into the preference learning framework. Through the recursive embedding propagation, both the short-term and long-term behavior-specific interaction patterns are preserved in the learned representations of users/items.

\end{itemize}

In the following subsections, we will elaborate the design motivations and technical details of each learning component in our proposed \model\ framework.

\subsection{Dynamic Individual Interest Modeling}
To capture the time-evolving user interests with the modeling of underlying relation heterogeneity across different types of behaviors, we develop a multi-behavior transformer network to distill short-term behavior dynamics. In particular, we explicitly model user short-term preference by splitting multi-behavior interaction sequence $S_i$ into $R$ fine-grained sub-sequences $S_i^r$ (indexed by $r$). Formally, we define $S_i^r = [(v_{1,i}^r, t_{1,i}^r, b_{1,i}^r),...,(v_{|S_i^r|,i}^r, t_{|S_i^r|,i}^r, b_{|S_i^r|,i}^r)]$. Without loss of generality, the individual interaction instance in the $r$-th sub-sequence is denoted as $(v_{k}^r, t_{k}^r, b_{k}^r)$.

\subsubsection{\bf Behavior-aware Context Embedding Layer} We design an embedding layer to jointly inject the multi-behavior contextual and temporal signals into the item representations. Following the mainstream recommendation paradigms~\cite{cao2019unifying,rendle2012bpr}, we first describe individual item $v$ with id-corresponding embedding $\textbf{e}_v \in \mathbb{R}^d$, where $d$ denotes the latent state dimensionality. Then, user behaviors are differentiated with behavior type embedding $\textbf{e}_b \in \mathbb{R}^d$ corresponding to $b$-th type of user-item interaction. Furthermore, to capture the behavior dynamics, we enhance the context embedding layer by introducing a temporal encoding strategy. Motivated by the positional embedding method in Transformer~\cite{vaswani2017attention}, we adopt the sinusoid functions over timestamp information to generate base time embeddings $\textbf{t}_k^r \in \mathbb{R}^{2d}$ with the following operations:
\begin{align}
    \textbf{t}_k^{r,(2l)}=\sin(\frac{\tau(t_k^r)}{10000^{\frac{2l}{d}}})/\sqrt{d} \\\nonumber ~~~~\textbf{t}_k^{r,(2l+1)}=\cos(\frac{\tau(t_k^r)}{10000^{\frac{2l+1}{d}}})/\sqrt{d}
\end{align}
\noindent where $\tau(\cdot)$ represents the time slot mapping function with temporal periods. The odd and even index of $2d$-dimensional vector $\textbf{t}_k^r$ are represented by $2l$ and $2l+1$. We further apply $\sqrt{d}$ as the scale factor to alleviate the effect of large embedding values. In addition, we introduce a tunable linear projection over the generated base time embedding $\textbf{t}_k^r$ to endow our model with learning flexibility of temporal context as $\textbf{e}_{t_k^r}=\textbf{t}_k^r\cdot\textbf{W}^t$, where $\textbf{W}^t\in\mathbb{R}^{2d\times d}$. After that, we inject the compound contextual signals of multi-behavioral patterns and interaction dynamics with the integration operation as: $\textbf{E}_{k}^r=\textbf{e}_{v_k^r} \oplus \textbf{e}_{b_k^r} \oplus \textbf{e}_{t_k^r}$, where $\textbf{E}_k^r\in\mathbb{R}^d$ is the learned context-aware item embedding, and $\oplus$ denotes the element-wise addition operation.

\subsubsection{\bf Item-wise Sequential Dependency}
To capture the dynamism of short-term transitions across different items, our item-wise sequential dependency encoder is built upon the transformer architecture, based on self-attention mechanism under multi-head representation spaces. In specific, the encoded embedding of $k$-th interacted item $v_k^r$ is calculated by the following multi-head dot-product attention for item-wise mutual relation learning:
\begin{align}
    \bar{\textbf{E}}_k^r = \mathop{\Bigm|\Bigm|}\limits_{h=1}^H \sum_{k'=1}^K \alpha_{k,k'}\textbf{V}^h\textbf{E}_{k'}^r \nonumber\\
    \alpha_{k,k'}=\sigma(\frac{(\textbf{Q}^h\textbf{E}_k^r)^\top (\textbf{K}^h\textbf{E}_{k'}^r)}{\sqrt{d/H}})
\end{align}
\noindent where $\alpha_{k,k'}$ is the learned item-wise relevance score between $v_k^r$ and $v_{k'}^r$. $\sigma(\cdot)$ denotes the softmax function. The item representation refinement ($\bar{\textbf{E}}_k^r$) is conducted through the above aggregation to capture relations across different items. We enable our encoder with multiple representation subspaces to perform sequential information aggregation from different semantic dimensions (\ie, $h\in H$ attention heads). $\Bigm|\Bigm|$ denotes the vector concatenation operator. $\textbf{Q}^h, \textbf{K}^h, \textbf{V}^h\in\mathbb{R}^{\frac{d}{H}\times d}$ denote the $h$-th head query, key and value transformations, respectively. With the integration of context embedding layer and item-wise dependency encoder, our \model\ not only preserves the cross-item transitional relations, but also captures the multi-behavioral dynamism.

\subsection{Multi-Behavior Pattern Aggregation}
In this section, we perform the multi-behavior pattern aggregation with the modeling of time-aware user-item interactions. Towards this end, a bipartite graph $G_r=\{u_i^r \cup S_{i,r}, \mathcal{E}_r\}$, where $u_i^r$ denotes the intermediate user vertex corresponding to the $r$-th interaction sub-sequence $S_{i,r}$. $\mathcal{E}_i^r$ represents the interactions between user $u_i$ and all items included in the sequence $S_{i,r}$. To capture the short-term behavior heterogeneity, we design a behavior-aware message passing scheme to differentiate and aggregate interactive patterns from different behavior types. This is a two-phase message passing paradigm which consists of propagating and refining embeddings from user to item side and vice versa. Given the learned item representations $\bar{\textbf{E}}_k^r\in\mathbb{R}^d$ by injecting multi-behavioral temporal context, the message passing from item to user side can be formally represented as follows:
\begin{align}
    \textbf{H}_b^r=\text{Aggre}(\bar{\textbf{E}}^r, b)=\sigma(\sum_{k=1}^K\phi(b_k^r=b)\cdot\bar{\textbf{E}}_k^r\cdot\textbf{W}_b)
\end{align}
\noindent where $\textbf{H}_b^r \in\mathbb{R}^d$ represents the generated embedding of node $u_i^r$ under the $b$-th behavior type. We utilize $\textbf{W}_b$ as behavior-aware transformation to enhance the modeling of behavior-specific semantics. $\sigma(\cdot)$ is the ReLU activation function. We define $\phi(\cdot)$ as an indicator function to fit the behavior type-aware embedding propagation, \ie, $\phi(\cdot)=1$ given $b_k^r=b$.

\subsubsection{\bf Multi-Channel Projection} To effectively discriminate multi-typed of interactions during the message passing process, we adopt a multi-channel parameter learning strategy for the calculation of transformation $\textbf{W}_b\in\mathbb{R}^{d\times d}$, which can be formalized as follows:
\begin{align}
    \textbf{W}_b=\sum_{h=1}^H\beta_h\cdot\bar{\textbf{W}}_h;~~~~~
    \beta_h=\sigma(\textbf{P}\cdot\textbf{b} + \mathbf{\mu})(h)
    \label{eq:multi_channel}
\end{align}
where $\bar{\textbf{W}}\in\mathbb{R}^{d\times d}$ denotes the $H$ base transformations (indexed by $h$) shared by all behavior types. $\beta_h$ is the learned weight for the $h$-th latent channel representation, which is calculated by a fully-connected layer with transformation $\textbf{P}\in\mathbb{R}^{H\times d}$ and bias $\mathbf{\mu}\in\mathbb{R}^{H}$. 

In our \model\ framework, the multi-channel projection layer serve as an important component in the global item-wise dependency modeling with the preservation of behavior heterogeneity. Specifically, our designed multi-channel projection aims to differentiate the multi-relational interactions during the message passing across time. After the behavior embedding projection over multiple base transformations, the behavior-aware semantics are encoded with the developed channel-wise aggregation layer, which endows our \model\ method to preserve the inherent behavior semantics of different types of user-item interactions.


\subsubsection{\bf Aggregation over Cross-Type Relations}
After encoding the type-specific behavior semantics, the next step is to feed the behavior-aware embeddings into an aggregation layer to uncover the implicit relations across different types of behavioral patterns. Different from static attentive relation aggregation schemes~\cite{xin2019relational,wang2020ckan}, we fit our learning scenario with the dynamic user preference, by proposing an adaptive attention network to distinguish influence of different behavior types in predicting the target one, with the awareness of time-evolving multi-behavioral interaction patterns.
\begin{align}
    \label{eq:userEmbed}
    \bar{\textbf{H}}^r &=\sum_{b=1}^B\gamma_b\cdot\textbf{H}_b^r \nonumber\\
    \gamma_b & =\sigma_1({\textbf{H}_b^r}^\top \cdot \sigma_2(\sum_{b=1}^B\textbf{H}_b^r\textbf{W}_A + \mathbf{\mu}_A)
\end{align}
where $\textbf{W}_A\in\mathbb{R}^{d\times d}, \mathbf{\mu}_A\in\mathbb{R}^d$ are the transformation and bias for our aggregation layer, respectively. $\sigma_1(\cdot)$ denotes the softmax function, and $\sigma_2(\cdot)$ denotes the ReLU function. $\gamma_b$ is the learned attentive weight for the $b$-th behavior type, calculated using the summation of all behavioral embeddings as query. As a result, we explicitly preserve the time-aware multi-behavior contextual information in the user representation $\bar{\textbf{H}}^r$.

To model time-aware user's influence over different items, we perform the message passing from user to item side through the aforementioned embedding propagation and aggregation schemes:
\begin{align}
    \bar{\textbf{E}}_{j}^b=\sigma(\sum_{v_k^r=j}\phi(b_k^r=b)\cdot\bar{\textbf{H}}^r\cdot\textbf{W}_b);~~~
    \bar{\textbf{E}}_j = \sum_{b=1}^B\gamma_b\cdot \bar{\textbf{E}}_{j,b}
\end{align}
\noindent where $\textbf{W}_b$ is calculated by the multi-channel projection scheme with different transformation mappings (Eq~\ref{eq:multi_channel}), and $\gamma_b$ is returned from our attentive aggregation mechanism. It is worth noting that $\bar{\textbf{E}}_{j}^b, \bar{\textbf{E}}_j \in\mathbb{R}^d$ are cross-time embeddings of item $v_j$ (without super-script $r$), which could capture changes in item semantics over time.

\subsection{Global Relational Context Learning}

Since user's preference is inherently dynamic and evolve over time, our \model\ further proposes to inject the long-range dynamism into the global-level graph relation encoder, which captures the dynamic multi-behavior patterns from both the long-term and short-term perspectives. We define a global user graph to contain all users ($u_i \in \mathcal{U}$) and their intermediate sub-sequence nodes ($u_i^r \in \mathcal{U}_r$), \ie, $G=\{\mathcal{U}, \mathcal{U}_r, \mathcal{E}\}$, where $\mathcal{E}$ represents the connections between each user $u_i$ and his/her intermediate sub-sequence nodes $u_i^r$, \ie, an edge connects $u_i$ and $u_i^r$ ($1\leq r \leq R$). In this component, the global user embedding $\bar{\mathbf{\Gamma}}_i \in\mathbb{R}^d$ is generated by performing graph-structured information aggregation, over short-term user preference embeddings ($\bar{\textbf{H}}^r$) across different interaction sub-sequences.


\subsubsection{\bf Global-level User Representation}
To learn global-level user representations, we incorporate the temporal context into the message passing procedure of attentional graph neural architecture, to allow the short-term preference encoded with different timestamps to interact with each other in a differentiable manner:
\begin{align}
    \bar{\mathbf{\Gamma}}_i &= \Phi(\bar{\textbf{H}}^r)=\sigma(\sum_{r=1}^{R_i}\eta_r \cdot \bar{\textbf{H}}^r) \nonumber\\
    \eta_r &= \mathbf{\Gamma}_i^\top \cdot \bar{\textbf{H}}^r;~ \bar{\textbf{H}}^r=\mathbf{\Gamma}_i+\textbf{t}^r
\end{align}
\noindent where $\eta_r$ is the learned attentive weights for pairwise relationship between $u_i$ and $u_i^r$. $\Phi(\cdot)$ denotes the global information aggregation function with the $\sigma(\cdot)$ activation function. $\textbf{t}^r\in\mathbb{R}^d$ is our temporal embedding encoded from our context embedding layer. The sub-sequence representation $\bar{\textbf{H}}^r$ will be updated with the injection of temporal localized information as: $\bar{\textbf{H}}^r=\mathbf{\Gamma}_i+\textbf{t}^r$, where $\mathbf{\Gamma}_i\in\mathbb{R}^d$ is the user embedding which is either initialized or smoothed by the neighboring nodes. By doing so, \model\ maintains the user preference embeddings from different time periods as a unified feature embedding for global user representation, with the modeling of temporal dependency and evolution of user interaction patterns.



\subsection{High-order Relation Aggregation}
As discussed above, \model\ performs two-phase information aggregation for encoding information of i) short-term multi-behavior interactions between user and item ; ii) long-range dynamic structural dependency of user interest across time durations. Generally, our \model\ architecture takes the relational structure of $G_r$ and $G$ as computation graphs for embedding propagation, during which the local relational features from neighborhood will be aggregated to obtain a contextual representation. Such message passing paradigm can be generalized with the incorporation of high-order connectivity as follows (suppose $l$ denotes the $(l)$-th GNN layer in the graph-based information propagation architecture):
\begin{align}
    \bar{\textbf{H}}^{r,(l)}&=\sum_{b=1}^B\gamma_b \cdot (\sigma(\sum_{k=1}^K\phi(b_k^r=b)\cdot\bar{\textbf{E}}_k^{r,(l)}\cdot\textbf{W}_b))\nonumber\\
    \mathbf{\Gamma}_i^{(l+1)}&=\sigma(\sum_{r=1}^{R_i}\eta_r \cdot \bar{\textbf{H}}^{r,(l)}),~~\bar{\textbf{H}}^{r,(l+1)}=\mathbf{\Gamma}_i^{(l)}+\textbf{t}^r\nonumber\\
    \bar{\textbf{E}}_j^{(l+1)}&=\sum_{b=1}^B\gamma_b\cdot \sigma(\sum_{v_k^r=j}\phi(b_k^r=b)\cdot\bar{\textbf{H}}^{r,(l+1)}\cdot\textbf{W}_b)
\end{align}
\noindent where $\bar{\textbf{E}}_k^{r,(0)}, \bar{\mathbf{\Gamma}}_i^{(0)}$ are initialized id-corresponding item and user embeddings, respectively. We endow \model\ with the ability of capturing high-order collaborative effects across different users/items, by extending our developed graph relation encoder from one layer to multiple layers (\ie, $L$). We finally generate user ($\tilde{\mathbf{\Gamma}}_i \in\mathbb{R}^d$), item ($\tilde{\textbf{E}}_j \in\mathbb{R}^d$) and intermediate sub-sequence node ($\tilde{\textbf{H}}^r \in\mathbb{R}^d$) embeddings across $L$ graph layers with the element-wise summation:
\begin{align}
    \tilde{\mathbf{\Gamma}}_i=\sum_{l=0}^L\mathbf{\Gamma}_i^{(l)},~~ \tilde{\textbf{H}}^r=\sum_{l=0}^L\textbf{H}^{r,(l)},~~
    \tilde{\textbf{E}}_j=\sum_{l=0}^L\textbf{E}_j^{(l)}
\end{align}

\subsection{Model Prediction and Optimization}
Following the same settings in sequential recommender systems~\cite{wang2020next}, to reflect the dynamic characteristics of user preference, we leverage the time-aware user representation $\tilde{\textbf{H}}^r$ to make forecasting on future user-item interactions. The probability of future interaction between user $u_i$ and item $v_j$ is estimated by $\text{Pr}_{i,j}=\textbf{z}^\top(\tilde{\textbf{H}}^{(-1)}_i \circ \tilde{\textbf{E}}_j)$, where $\textbf{z}\in\mathbb{R}^d$ is a parametric vector. $\tilde{\textbf{H}}_i^{-1}$ represents the embedding of the last interacted item of user $u_i$ in the $r$-th interaction sub-sequence. $\circ$ denotes the element-wise multiplication operation. Next, we will describe our defined optimized objective and perform the time complexity analysis for our \model\ framework.

\subsubsection{\bf Optimization Objective}
During the training phase, the parameter inference process is optimized by leveraging historical user-item interactions for label data augmentation~\cite{kang2018self,sun2019bert4rec}. In this case, the prediction score for $u_i$ and $v_j$ during the time interval of sub-sequence $S_i$ is calculated as: $\text{Pr}_{i,j}^r=\textbf{z}^\top(\tilde{\textbf{H}}^{r}_i \circ \tilde{\textbf{E}}_j)$. We formally define our objective function with the marginal pairwise loss as follows:
\begin{align}
    \mathcal{L}=\sum_{i=1}^N\sum_{r=1}^{R_i}\sum_{c=1}^C\max(0,1-\text{Pr}_{i,p_c}^r+\text{Pr}_{i,n_c}^r)+\lambda\cdot\|\mathbf{\Theta}\|_\text{F}^2
\end{align}
\noindent where $N$ represents the number of users and $R_i$ denotes the number of interaction sub-sequences for user $u_i$. $p_c$ and $n_c$ refer to $C$ positive and the negative samples (indexed by $c$), respectively. We further apply the weight-decay is the weight-decay regularization term $\lambda\cdot\|\mathbf{\Theta}\|_\text{F}^2$ for alleviating overfitting phenomenon.

We summarize the learning process of our \emph{\model} in Alg~\ref{alg:learn_alg}. In particular, the model optimization involves several key steps: (i) We first generate behavior-ware context embeddings $\textbf{E}_k^r$. Then, we feed it into the item-wise sequential dependency encoder to obtain $\bar{\textbf{E}}_k^r$ corresponding to the $k$-th behavior type and $r$-th interaction sub-sequence. (ii) Through the recursive message passing procedure, we learn the sub-sequence behavior representation $\textbf{H}_b^r$ of users and inject the cross-type behavior relations into the refined embedding $\bar{\textbf{H}}^r$. (iii) The aggregated global user embedding $\bar{\mathbf{\Gamma}}_i$ preserves his/her long-term multi-behavior patterns. The sub-user and item embeddings are updated accordingly. (iv) Layer-specific embeddings are combined to generate cross-layer representations $\tilde{\mathbf{\Gamma}}_i, \tilde{\textbf{H}}^r, \tilde{\textbf{E}}_j$.

\begin{algorithm}[t]
\footnotesize
	\caption{Model Inference of \model\ Framework}
	\label{alg:learn_alg}
	\LinesNumbered
	\KwIn{user-item interaction sequences $S_i$, sub-user size $|S_i^r|$, graph iterations $L$, target behavior $B$, sample number $C$, maximum epoch number $E$, regularization weight $\lambda$, learning rate $\rho$}
	\KwOut{trained parameters in $\mathbf{\Theta}$}
	Initialize all parameters in $\mathbf{\Theta}$\\
    Divide $S_i$ into sub-sequences $S_i^r$ of size $|S_i^r|$\\
    \For{$e=1$ to $E$}{
        Generate behavior-aware context embeddings $\textbf{E}_k^r$\\
        Capture item-wise sequential dependency to yield $\bar{\textbf{E}}_k^r$\\
        \For{$l=1$ to $L$}{
            Compute sub-users' behavior embeddings $\textbf{H}_b^r$\\
            Conduct cross-type aggregation to get $\bar{\textbf{H}}^r$\\
            Aggregate for the global user embedding $\bar{\mathbf{\Gamma}}_i$\\
            Refine the sub-user embeddings $\bar{\textbf{H}}^r$ using $\bar{\mathbf{\Gamma}}_i$\\
            Update the item embeddings $\bar{\textbf{E}}_j$ using $\bar{\textbf{H}}^r$\\
        }
        Integrate the cross-order graph embeddings $\tilde{\mathbf{\Gamma}}_i, \tilde{\textbf{H}}^r, \tilde{\textbf{E}}_j$\\
        
        Draw a mini-batch $\textbf{U}$ from all users, each with $C$ positive-negative samples\\
        $\mathcal{L} = \lambda\cdot\|\mathbf{\Theta}\|_\text{F}^2$\\
        \For{each $u_i\in\textbf{U}$}{
            Compute predictions $\text{Pr}_{i,p_c}, \text{Pr}_{i,n_c}$\\
            $\mathcal{L}+=\sum_{c=1}^C\max(0,1-\text{Pr}_{i,p_c}+\text{Pr}_{i,n_c})$\\
        }
        \For{each parameter $\theta\in\mathbf{\Theta}$}{
            $\theta=\theta-\rho \cdot \partial \mathcal{L} / \partial \theta$
        }
    }
    return all parameters $\Theta$
\end{algorithm}

\subsubsection{\bf Model Complexity Analysis}
In this subsection, we analyze the time complexity of our \model\ model. In particular, in the encoding process of item-wise sequential dependency, the computational cost for pairwise item relation learning and embedding projection are $O(|\mathcal{E}_r|\times |S_i^r| \times d)$ and $O(|\mathcal{E}_r|\times d^2)$, respectively. Different from directly performing sequential learning over the entire item sequence $S_i$, our method significantly reduces the computational cost for self-attention operations. Additionally, during the message passing and aggregation in the graph neural architecture, \model\ costs $O(L\times|\mathcal{E}_r|\times d + (M+|\mathcal{E}|/{|S_i^r|})\times L\times d^2)$ for encoding the short-term multi-behavioral patterns, and $O(|\mathcal{E}|/S_i\times L\times d)$ for learning global relation context, where $L$ is the depth of our graph neural network. Considering the small value of $L$, this complexity term has very small influence on the overall model cost. Under the same experimental settings (\eg, sequence length and latent state dimensionality), \model\ could achieve comparable complexity to the most recently developed baselines: HyRec~\cite{wang2020next} and GCGNN~\cite{wang2020global}.

\section{Evaluation}
\label{sec:eval}

In this section, we perform experiments to validate the effectiveness of our proposed \emph{\model} recommendation framework by investigating the research questions as follows:

\begin{itemize}[leftmargin=*]
\item \textbf{RQ1}: How does our proposed \emph{\model} method perform compared with state-of-the-art recommendation methods? \\\vspace{-0.1in}

\item \textbf{RQ2}: How do the different components (\eg, behavior-aware transformer network and global relational context learning) of \emph{\model} contribute to the model performance? \\\vspace{-0.1in}

\item \textbf{RQ3}: How do the integration of different types of behavioral patterns affect the prediction of target behavior? \\\vspace{-0.1in}



\item \textbf{RQ4}: How do the configurations of model hyperparameters affect the recommendation performance? \\\vspace{-0.1in}


\item \textbf{RQ5}: Can the proposed \emph{\model} framework show model interpretability, through the effective modeling of sequential multi-behaviour patterns?

\end{itemize}


\begin{table}[t]
    \caption{Statistics of experimented datasets}
    \vspace{-0.1in}
    \label{tab:data}
    \centering
    \small
	\setlength{\tabcolsep}{0.6mm}
    \begin{tabular}{ccccc}
        \midrule
        Dataset& \# of Users & \# of Items & \# of Interactions \\
        \hline
        Taobao& 147894 & 99037 & 7658926 \\
        IJCAI& 423423 & 874328 & 36203512 \\
        \hline
    \end{tabular}
\vspace{-0.1in}
\end{table}

\subsection{Experiment Settings}
In this section, we first describe the details of our experimented datasets and introduce the evaluation metrics. Then, we present the baseline methods from a variety of research lines for performance comparison, as well as the parameter settings of our developed \emph{\model} framework.

\subsubsection{\bf Data Description}
\label{sec:data}

In our evaluation, experiments are performed on two public datasets (see Table~\ref{tab:data}) collected from real-world online platforms. Both datasets contain different types of user behaviors in e-commerce services. \textbf{Taobao-Data}: This data records four types of user-item interactions collected from the Taobao online retail system-one of the world's e-commerce marketplace. Particularly, multi-typed behaviors include \emph{page view}, \emph{add-to-cart}, \emph{tag-as-favorite} and \emph{purchase}. \textbf{IJCAI-Contest Data}: This data is released by IJCAI competition to record users' online activities from online business-to-consumer e-commerce site, also involving four types of behaviors: \emph{clicking}, \emph{adding-to-cart}, \emph{tagging with favor} and \emph{purchasing}. The temporally-ordered multi-typed interaction logs are used to generated multi-behavior sequence $S_i$ for each user ($u_i \in \mathcal{U}$). Because of the importance of customer purchase activities in practical e-commerce platforms~\cite{2019online,wu2018turning}, we consider the purchases of users as the \emph{target behaviors} and other types of user behaviors are regarded as the \emph{context behaviors} for representation enhancement.




\subsubsection{\bf Evaluation Metrics}
In our experiments, we evaluate the recommendation performance of all compared approaches using two widely adopted metrics: NDCG@N and Recall@N, so as to measure the accuracy of top-$N$ recommended item for different users. Note that the better recommendation performance is reflected by higher NDCG@N and Recall@N scores. Here, NDCG score reflects the position-aware recommendation accuracy which assigns high weights for higher ranked item hits. We utilize the leave-one-out evaluation~\cite{kang2018self,sun2019bert4rec} for generating the training and testing data instances based on the temporal information of user behavior data. Specifically, the last interacted item of each evaluated user is considered as the test instance for making recommendation.


\begin{table*}[t]
\caption{Overall performance comparison of all methods in terms of \textit{HR@$N$} and \textit{NDCG@$N$} ($N=1,5,10$).}
\centering
\scriptsize
\setlength{\tabcolsep}{1.3mm}
\begin{tabular}{|c|c|c|c|c|c|c|c|c|c|c|c|c|c|c|c|c|c|c|c|c|}
\hline
\multirow{3}{*}{Methods} & \multicolumn{10}{c|}{Taobao Data} & \multicolumn{10}{c|}{IJCAI Contest}\\
\cline{2-21}
& \multicolumn{2}{c|}{Top @ 1} & \multicolumn{4}{c|}{Top @ 5} & \multicolumn{4}{c|}{Top @ 10} & \multicolumn{2}{c|}{Top @ 1} & \multicolumn{4}{c|}{Top @ 5} & \multicolumn{4}{c|}{Top @ 10}\\
\cline{2-21}
& HR & Imp & HR & Imp & NDCG & Imp & HR & Imp & NDCG & Imp & HR & Imp & HR & Imp & NDCG & Imp & HR & Imp & NDCG & Imp\\
\hline
\hline

BPR & 0.048 & 139\% & 0.209 & 39\% & 0.143 & 26\% & 0.295 & 53\% & 0.179 & 47\% & 0.073 & 102\% & 0.184 & 117\% & 0.130 & 125\% & 0.257 & 102\% & 0.153 & 116\% \\
\hline
NCF & 0.061 & 89\% & 0.231 & 26\% & 0.157 & 15\% & 0.325 & 39\% & 0.201 & 31\% & 0.139 & 6\% &  0.351 & 14\% & 0.255 & 15\% & 0.459 & 13\% & 0.294 & 12\% \\
\hline
DeepFM & 0.059 & 95\% &  0.223 & 30\% & 0.142 & 27\% & 0.328 & 38\% & 0.175 & 50\% & 0.138 & 7\% & 0.332 & 20\% & 0.244 & 20\% & 0.469 & 11\% & 0.290 & 14\%\\
\hline
GRURec & 0.081 & 42\% & 0.237 & 22\% & 0.161 & 12\% & 0.328 & 38\% & 0.187 & 41\% & 0.107 & 38\% & 0.284 & 41\% & 0.194 & 51\% & 0.418 & 24\% & 0.247 & 34\% \\
\hline
Caser & 0.094 & 22\% &  0.239 & 21\% & 0.168 & 7\% & 0.364 & 24\% & 0.197 & 34\% & 0.129 & 14\% &  0.295 & 35\% & 0.208 & 41\% & 0.422 & 23\% & 0.252 & 31\% \\
\hline
NARM & 0.085 & 35\% & 0.220 & 32\% & 0.151 & 19\% & 0.320 & 41\% & 0.181 & 45\% & 0.107 & 38\% & 0.289 & 38\% & 0.200 & 47\% & 0.396 & 31\% & 0.229 & 44\% \\
\hline
SASRec & 0.089 & 29\% &  0.250 & 16\% & 0.169 & 7\% & 0.353 & 28\% & 0.203 & 30\% & 0.114 & 29\% &  0.306 & 30\% & 0.216 & 36\% & 0.420 & 24\% & 0.247 & 34\% \\
\hline
Bert4Rec & 0.113 & 2\% &  0.265 & 9\% & 0.174 & 3\% & 0.369 & 22.5\% & 0.229 & 15\% & 0.141 & 4\% &  0.356 & 12\% & 0.261 & 12\% & 0.467 & 11\% & 0.297 & 11\% \\
\hline
Chorus & 0.100 & 15\% & 0.235 & 23\% & 0.171 & 5\% & 0.388 & 17\% & 0.232 & 13\% & 0.140 & 5\% & 0.345 & 15.7\% & 0.247 & 19\% & 0.457 & 14\% & 0.283 & 17\% \\
\hline
HyRec & 0.088 & 31\% & 0.228 & 27\% & 0.161 & 12\% & 0.324 & 40\% & 0.191 & 38\% & 0.137 & 8\% &  0.323 & 24\% & 0.229 & 28\% & 0.442 & 17\% & 0.266 & 24\% \\
\hline
MAGNN & 0.108 & 6\% &  0.235 & 23\% & 0.174 & 3\% & 0.320 & 41\% & 0.199 & 32\% & 0.127 & 16\% &  0.291 & 37\% & 0.212 & 38\% & 0.392 & 32\% & 0.245 & 35\% \\
\hline
SR-GNN & 0.086 & 34\% & 0.240 & 21\% & 0.162 & 11\% & 0.332 & 36\% & 0.191 & 38\% & 0.133 & 11\% & 0.316 & 26\% & 0.214 & 37\% & 0.431 & 20\% & 0.263 & 26\% \\
\hline
GCGNN & 0.087 & 32\% &  0.250 & 16\% & 0.168 & 7\% & 0.349 & 30\% & 0.201 & 31\% & 0.136 & 8\% &  0.289 & 38\% & 0.210 & 40\% & 0.428 & 21\% & 0.257 & 28\% \\
\hline
MGNN & 0.096 & 20\% & 0.259 & 12\% & 0.172 & 5\% & 0.374 & 21\% & 0.219 & 20\% & 0.138 & 7\% &  0.244 & 64\% & 0.248 & 18\% & 0.457 & 14\% & 0.286 & 15\% \\
\hline
HGT & 0.101 & 14\% & 0.261 & 11\% & 0.174 & 3\% & 0.364 & 24\% & 0.216 & 22\% & 0.128 & 15\% &  0.309 & 29\% & 0.207 & 41.5\% & 0.450 & 15\% & 0.265 & 25\% \\
\hline
NMTR & 0.079 & 46\% & 0.218 & 33\% & 0.147 & 22\% & 0.332 & 36\% & 0.179 & 47\% & 0.141 & 4\% &  0.360 & 10.8\% & 0.254 & 15\% & 0.481 & 8\% & 0.304 & 9\% \\
\hline
MATN & 0.081 & 42\% & 0.226 & 29\% & 0.153 & 17\% & 0.354 & 28\% & 0.209 & 26\% & 0.142 & 4\% &  0.375 & 6\% & 0.273 & 7\% & 0.489 & 6\% & 0.309 & 7\% \\
\hline
MBGCN & 0.110 & 5\% & 0.259 & 12\% & 0.172 & 5\% & 0.369 & 23\% & 0.222 & 19\% & 0.137 & 8\% &  0.332 & 20\% & 0.228 & 31\% & 0.463 & 12\% & 0.277 & 19\% \\
\hline
\hline
\emph{\model} & \textbf{0.115} & -- &  \textbf{0.290} & -- & \textbf{0.180} & -- & \textbf{0.452} & -- & \textbf{0.263} & -- & \textbf{0.148} & & \textbf{0.399} & -- & \textbf{0.293} & -- & \textbf{0.519} & -- & \textbf{0.330} & -- \\
\hline

\end{tabular}
\vspace{-0.05in}
\label{tab:target_behavior}
\end{table*}

\subsubsection{\bf Methods for Comparison}
To comprehensively evaluate the performance, we compare our \emph{\model} with state-of-the-art recommendation techniques which can be grouped into different model types: \\\vspace{-0.1in}

\noindent \textbf{Conventional Matrix Factorization Model}: We first consider the representative matrix factorization method--Bayesian personalized ranking as the baseline.
\begin{itemize}[leftmargin=*]
\item \textbf{BPR}~\cite{rendle2012bpr}. It enhances the matrix factorization model with pairwise loss for learning personalized rankings.\\\vspace{-0.1in}
\end{itemize}

\noindent \textbf{Neural Collaborative Filtering Methods}: The conventional collaborative filtering methods have been enhanced by deep learning techniques for nonlinear deep feature extraction.
\begin{itemize}[leftmargin=*]
\item \textbf{NCF}~\cite{he2017neural}. This approach replaces the dot-product operation in matrix factorization with multi-layer perceptrons to model the non-linearity in user-item interactions. \\\vspace{-0.12in}
\item \textbf{DeepFM}~\cite{guo2017deepfm}. This wide-and-deep model augments classic factorization machine with deep neural networks, to consider both low- and high-order feature interactions.
\end{itemize}

\noindent \textbf{RNN-based Sequential Recommendation Method}: Recurrent neural network has been utilized in sequential recommendation model for dynamic behavior modeling.
\begin{itemize}[leftmargin=*]
\item \textbf{GRURec}~\cite{hidasi2015session}. This method adopts the gated recurrent unit to model sequential behaviors of users, based on the gating mechanism to control the information flow.\\\vspace{-0.1in}
\end{itemize}

\noindent \textbf{Convolution-based for Sequential Recommendation}: We consider the sequential recommendation method using convolutional filters to encode local temporal patterns.
\begin{itemize}[leftmargin=*]
\item \textbf{Caser}~\cite{tang2018personalized}. It utilizes the convolution neural network to perform information aggregation across temporally-order items in both horizontal and vertical
way.\\\vspace{-0.1in}
\end{itemize}

\noindent \textbf{Attention/Transformer-based Neural Network for Sequential Recommender Systems}: Inspired by the effectiveness of attention mechanism in relational learning, there exist many attentional recommendation frameworks for modeling user dynamic preference in sequential recommendation.
\begin{itemize}[leftmargin=*]
\item \textbf{NARM}~\cite{li2017neural}. It jointly learns the sequential patterns of user interactions as well as the session-specific general interest with the integration of recurrent network and attention mechanism. \\\vspace{-0.1in}

\item \textbf{SASRec}~\cite{kang2018self}. In SASRec, self-attention is employed to encode the sequential patterns of user interaction, without the recurrent operations over input sequences. The embedding transformation is performed with query and key dimensions. \\\vspace{-0.1in}

\item \textbf{Bert4Rec}~\cite{sun2019bert4rec}. It is a transformer-based sequential recommender system with bidirectional self-attention architecture and cloze objective to incorporate sequential signals.\\\vspace{-0.1in}
\end{itemize}

\noindent \textbf{Knowledge-aware Sequential Recommendation}: We also include the sequential recommendation baseline with the consideration of knowledge-aware item relationships.
\begin{itemize}[leftmargin=*]
\item \textbf{Chorus}~\cite{chorus2020}. This method accounts for both temporal context and item correlations in the sequential recommendation model. For this method, the TransE is adopted as the translation function to obtain entity representations. \\\vspace{-0.1in}
\end{itemize}

\noindent \textbf{Sequential/Session-based Recommendation Models with Graph-based Neural Networks}: Another important research line of time-aware recommender systems lies in the utilization of graph neural networks for behavior dependence learning.
\begin{itemize}[leftmargin=*]

\item \textbf{HyRec}~\cite{wang2020next}. It predicts the next interacted item of users by utilizing the hypergraph neural network to model the short-term item relations. Then, a fusion layer is introduced to aggregate dynamic item semantics. \\\vspace{-0.1in}

\item \textbf{MAGNN}~\cite{ma2020aaai}. It captures both the short- and long-term user preference with memory-augmented graph neural network. The generated short- and long-term pattern embeddings are aggregated through an interesst fusion layer. \\\vspace{-0.1in}

\item \textbf{MGNN}~\cite{beyongwang}. It is a relation-aware graph neural network-based architecture which utilizes the gating mechanism for representation fusion, with the consideration of multi-relational user sequence. \\\vspace{-0.1in}

\item \textbf{SR-GNN}~\cite{wu2019session}. It designs a gated graph neural network to generate session-level item representations with transition regularities. The constructed graph data contains different session sequences in session-based recommendation. \\\vspace{-0.1in}

\item \textbf{GCGNN}~\cite{wang2020global}. This approach performs the information aggregation over the constructed session graphs by considering pairwise item-transition information. \\\vspace{-0.1in}

\end{itemize}

\noindent \textbf{Heterogeneous Graph Neural Model}: Additionally, we consider another type of baseline by utilizing the heterogeneous graph neural network to capture the behavior heterogeneity during the information propagation procedure.
\begin{itemize}[leftmargin=*]
\item \textbf{HGT}~\cite{heterogeneous2020}. It is a state-of-the-art heterogeneous graph learning model which integrates the transformer with the graph-based message passing scheme. We utilize its heterogeneous graph relation encoder to model different types of user-item interactions. \\\vspace{-0.1in}
\end{itemize}

\noindent \textbf{Multi-Behavior Recommendation Frameworks}: Several recent studies attempt to capture multi-behavioral context for recommendation by encoding behavior dependency based on various techniques.
\begin{itemize}[leftmargin=*]

\item \textbf{NMTR}~\cite{gao2019neural}: It integrates the neural collaborative filtering and multi-task learning paradigm to model the behavior-wise dependency of users with pre-defined correlations. \\\vspace{-0.1in}

\item \textbf{MATN}~\cite{xia2020multiplex}: This recommendation method designs memory-based neural units to enhance the attention mechanism, to capture the behaviour patterns of users. It only encodes the local connectivity patterns between users and items and cannot capture the high-order dependence. \\\vspace{-0.1in}

\item \textbf{MBGCN}~\cite{jin2020multi}: it is a state-of-the-art graph-based recommendation model which alleviates the data sparsity issue with the modeling of multi-typed user behavior data. In this model, the graph convolution operations are conducted to perform message passing.

\end{itemize}


\subsubsection{\bf Parameter Settings}
We implement our \emph{\model} using TensorFlow and adopt Adam for model optimization. During the training phase, the model inference process is conducted with the learning rate of $1e^{-3}$ (configured with 0.96 decay rate). The batch size is selected from the range $\{32, 128, 256, 512\}$ with the best setting of 256 for Taobao dataset and 512 for IJCAI dataset. The weight-decay regularization term is configured with $\lambda$ selected from \{0.05, 0.01, 0.005, 0.001, 0.0005\}. The default settings for hidden state dimensionality is 16. The length of user's sub-sequence is chosen from the range $[2,4,6,8,10]$ with the best setting of 6 for Taobao dataset and 10 for IJCAI dataset. We configure our self-attention mechanism under 2 heads of representation subspaces. The number of channels for base transformation in behavior semantic encoding is set as 2. The number of information propagation layers $L$ in our \emph{\model} is chosen from $[1,2,3]$. For a fair comparison settings, all compared neural network models are evaluated using their released source code or implementing them according to their original papers. Detailed model hyperparameter settings can also be found in our released model implementation in the abstract section.


\subsection{Performance Comparison (RQ1)}
We present the performance comparison results of our \emph{\model} and baselines in Table~\ref{tab:target_behavior} in terms of (HR@1), (HR@5, NDCG@5) and (HR@10, NDCG@10). As we can see, \emph{\model} consistently outperforms all compared methods by a large margin under all metrics. We attribute such improvements to the multi-behavior dependency modeling of \emph{\model}: i) through uncovering multi-behavior user intentions, \emph{\model} can better characterize the user-item relations; ii) benefiting from our designed temporal graph neural transformer architecture, both short- and long-term behavior-type dependent interests of users are effectively preserved in our learned user and item representations.

We further summarize several observations as follow:

\begin{itemize}[leftmargin=*]

\item (1) Our \model\ method consistently outperforms baesline methods BPR, NCF, DeepFM in all cases on different datasets, which validates the superiority of our proposed multi-behavior sequential recommender. Most of sequential recommendation methods achieves better performance than BPR, NCF and DeepFM. This observation suggests the benefits of modeling behavior dynamics with sequential interaction patterns for predicting user preference. Furthermore, the performance superiority of multi-behavior recommender systems as compared to NCF and DeepFM justifies the importance of incorporating multi-behavior dependencies into the modeling of complex and diverse user preferences.\\\vspace{-0.1in}

\end{itemize}

\begin{itemize}[leftmargin=*]
\item (2) Bert4Rec method achieves better performance than other attention-based approaches (\ie, NARM, SASRec) and convolution-based sequential recommender system (\ie, Caser), which indicates the effectiveness of transformer in encoding sequential patterns. However, the performance improvement of our method over Bert4Rec shows that \emph{\model} significantly boosts the recommendation performance by embracing sequential multi-behavior modeling for recommendation. \\\vspace{-0.1in}
\end{itemize}

\begin{itemize}[leftmargin=*]
\item (3) The performance gap between \emph{\model} and GNN-based recommender systems with the modeling of temporal information (\ie, HyRec, MAGNN, SR-GNN, GCGNN), clarifying the importance of encoding behavior heterogeneity during the message passing paradigms. In particular, our \emph{\model} is equipped with dynamic multi-behavior modeling under a graph-based global relation aggregation, which shows the superior performance in encoding users' diverse preference. \\\vspace{-0.1in}
\end{itemize}

\begin{itemize}[leftmargin=*]
\item (4) For a comprehensive performance comparison settings, we also compare \emph{\model} with two recently developed time-aware recommendation (MAGNN) and network embedding (HGT) methods with heterogeneous graph neural networks. As relation-aware graph neural models, these models are generally superior to other GNN-based methods by considering the interaction heterogeneity. However, they fail to capture dynamic cross-type behavior dependencies from both short- and long-term perspectives, which limits their capability in effectively transferring knowledge among different types of user behaviors. \\\vspace{-0.1in}
\end{itemize}

\begin{itemize}[leftmargin=*]
\item (5) As Table~\ref{tab:target_behavior} shows, \emph{\model} effectively improves the recommendation performance when competing with state-of-the-art multi-behavior recommendation models (\ie, NMTR, MBGCN and MATN). We can observe the relative improvement ratio (measured by NDCG@10) between our \emph{\model} and MBGCN is 22.5\% and 12.1\% on Taobao and IJCAI-Contest dataset, respectively. Such performance improvement further verifies the rationality of our designed temporal graph neural architecture, which enables \emph{\model} to accurately capture the dynamic multi-behavior patterns. In most existing multi-behavior recommender systems, the time-evolving cross-type behavior dependencies are ignored, which degrades the performance.
\end{itemize}

\begin{table}[t]
\vspace{-0.05in}
\caption{Ablation study on IJCAI Contest dataset.}
\centering
\begin{tabular}{c|cc|cc}
\hline
Setting & \multicolumn{2}{c|}{Top-5} & \multicolumn{2}{c}{Top-10}\\
\hline
Metrics & HR & NDCG & HR & NDCG \\
\hline
w/o CE & 0.392 & 0.290 & 0.515 & 0.321 \\
w/o SD & 0.367 & 0.265 & 0.499 &0.306 \\
w/o MCP & 0.346 & 0.240 & 0.490 & 0.306  \\
w/o LBD & 0.387 & 0.283 & 0.511 & 0.319\\
w/o CTA & 0.392 & 0.287 & 0.510 & 0.323 \\
FBA & 0.379 & 0.274 & 0.511 & 0.320\\
\hline
\emph{\model} & \textbf{0.399} & \textbf{0.293} & \textbf{0.519}  & \textbf{0.330}\\
\hline
\hline
Setting & \multicolumn{2}{c|}{Top-15} & \multicolumn{2}{c}{Top-20}\\
\hline
Metrics & HR & NDCG  & HR  & NDCG  \\
w/o CE & 0.603 & 0.354 & 0.670 & 0.367\\
w/o SD & 0.591 & 0.332 & 0.661 & 0.349 \\
w/o MCP & 0.575 & 0.308 & 0.647 & 0.325 \\
w/o LBD & 0.597 & 0.348 & 0.658 & 0.364 \\
w/o CTA & 0.603 & 0.352 & 0.666 & 0.368\\
FBA & 0.604 & 0.343 & 0.677 & 0.359\\
\hline
\emph{\model} & \textbf{0.608} & \textbf{0.356}  & \textbf{0.681} & \textbf{0.374} \\
\hline
\end{tabular}
\vspace{-0.05in}
\label{tab:module_ablation}
\end{table}

\subsection{Model Ablation Study (RQ2)}
To verify the effects of the designed different components in our model, we perform a detailed model ablation study of \emph{\model} on the IJCAI-Contest dataset. Specifically, we design different variants to make comparison with our \emph{\model} recommendation framework:


\begin{itemize}[leftmargin=*]

\item \textbf{\emph{w/o CE (Context Emebedding)}}: The time- and behavior-aware context are ignored during the sequential dependency encoding. Only item and positional embeddings are fed into the sequential dependency encoding component. \\\vspace{-0.1in}

\item \textbf{\emph{w/o SD (Sequential Dependency)}}: We do not involve the multi-behavior transformer for capturing short-term preference, and directly apply the temporal graph neural network to model the behavior heterogeneity. \\\vspace{-0.1in}

\item \textbf{\emph{w/o MCP (Multi-Channel Projection)}}: This variant simplifies the designed multi-behavior message passing paradigm without the multi-channel projection layer to preserve distinct behavior semantics. \\\vspace{-0.1in}

\item \textbf{\emph{w/o LBD (Long-Term Behavior Dynamics)}}: This model does not take the long-term multi-behavior dynamics into consideration. It ignores the global relation learning with behavior type-aware message passing component. \\\vspace{-0.1in}

\item \textbf{\emph{w/o CTA (Cross-Type Behavior Aggregation)}}: This model variant does not explore the cross-type behavioral relations with our attentive aggregation layer. Instead, we directly concatenate the propagated type-specific behavior embeddings to generate the user representation.

\end{itemize}

\noindent We show the ablation results in Table~\ref{tab:module_ablation} and discuss the findings and summarization from different aspects as below: \\\vspace{-0.12in}

\noindent (1) \textbf{Effect of behavior-aware context embedding}. Compared to w/o CE variant, \emph{\model} achieves better results. We owe it to the injection of multi-behavioral context into the modeling of item-wise sequential dependency, which allows the model to preserve short-term multi-behavior semantics. \\\vspace{-0.1in}

\noindent (2) \textbf{Effect of item-wise short-term dependency modeling}. Our multi-behavior transformer framework enhances the valina transformer with the incorporation of behavior-aware context into the modeling process of item-wise sequential dependencies. By comparing the performance of \emph{\model} with w/o SD, it is clear that with the multi-behavior transformer network, \emph{\model} achieves better performance by exploring short-term item-transition information under multi-behavior context. The designed multi-behavior transformer enables the item sequence encoder with the awareness of multi-behavior contextual signals between users and items, so as to capture the heterogeneous item-wise multi-relational dependencies. \\\vspace{-0.1in}

\noindent (3) \textbf{Effect of multi-channel projection}. We can observe that \emph{\model} is superior to w/o LBD. This demonstrates that the multi-channel projection layer is beneficial to enhance the encoding of behavior semantics during the multi-behavior pattern aggregation. \\\vspace{-0.1in}

\noindent (4) \textbf{Effect of long-term multi-behavior pattern aggregation}. The performance gap between \emph{\model} and w/o LBD indicates the importance of capturing user's long-term preference through discriminating different types of interactions. To characterize multi-behavior sequential patterns, we design the graph-structured embedding propagation to learn global-level user/item representations. \\\vspace{-0.1in}

\noindent (5) \textbf{Effect of cross-type behavior dependency learning}. With the designed attention-based aggregation network, the recommendation performance of our \emph{\model} is improved over the variant w/o CTA (performing simplified embedding concatenation). In the variant w/o CTA, our graph learning framework will give same weights to the propagated type-specific behavior representations, which may weaken the multi-behavior collaborative effect modeling. This suggests that the implicit relations across different types of interaction behaviors cannot be equally weighted. In addition, we design a new method variant FBA which fuses the multi-behavior patterns with the behavior-specific interaction frequency information. From the results, we can observe that our \model\ consistently outperforms the FBA variant with different top-N item positions (\eg, top-5, top-10, top-20). This further justifies the effectiveness of the learnable weight mechanism for cross-type behavior dependency modeling. \\\vspace{-0.1in}

\subsection{Effects of Context Behaviors (RQ3)}
This section evaluates the influence of type-specific behavior data in the recommendation performance for users' target behaviors. For IJCAI-Contest data, we define ``-pv'', ``-fav'', ``-cart'' to represent our multi-behavior sequential recommendation framework by removing the individual behavior type of ``page view'', ``tag-as-favorite'', ``add-to-cart'' behavior, respectively. In our ablation study, another variant ``+buy'' is designed to only contain the target type of user behaviors for making recommendation. \emph{\model} denotes the full version of our method to incorporate all types of user behaviors.\\\vspace{-0.1in}

\begin{figure}[h]
	\centering
	\subfigure[][IJCAI-HR]{
		\centering
		\includegraphics[width=0.37\columnwidth]{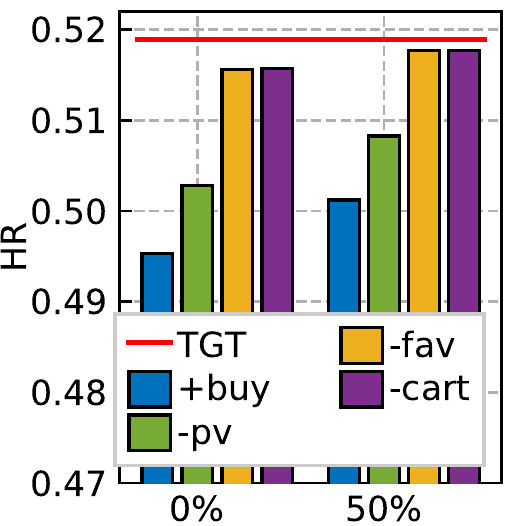}
		\label{fig:ab_ijcai_hr}
	}
	\subfigure[][IJCAI-NDCG]{
		\centering
		\includegraphics[width=0.37\columnwidth]{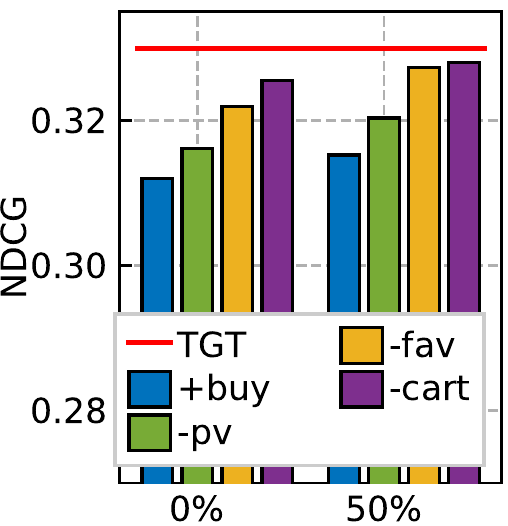}
		\label{fig:ab_ijcai_ndcg}
	}
	\subfigure[][Tmall-HR]{
		\centering
		\includegraphics[width=0.37\columnwidth]{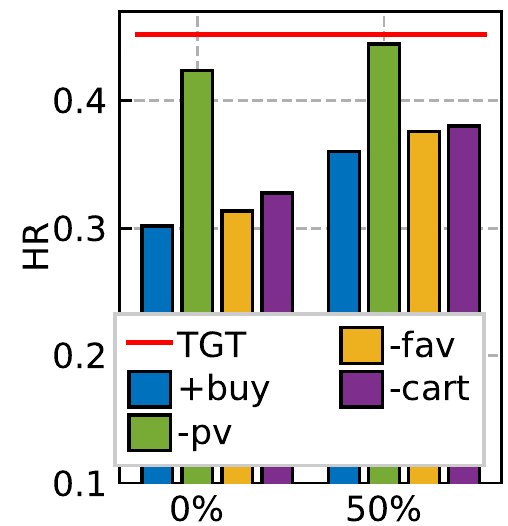}
		\label{fig:ab_tmall_hr}
	}
	\subfigure[][Tmall-NDCG]{
		\centering
		\includegraphics[width=0.37\columnwidth]{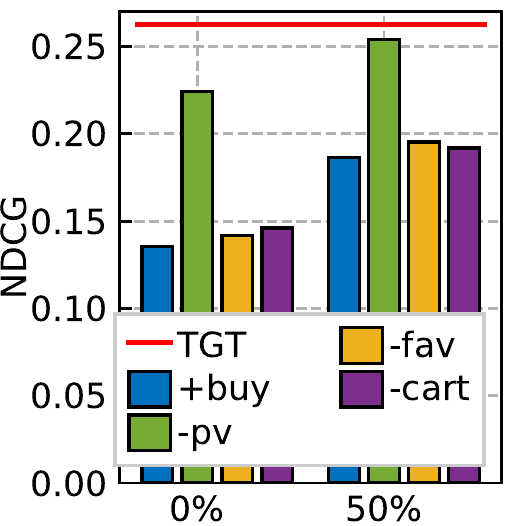}
		\label{fig:ab_tmall_ndcg}
	}
	\caption{Impact study of behavioral relation integration.}
	\label{fig:beh_ablation}
	\vspace{-0.1in}
\end{figure}

The results (measured by HR@10 and NDCG@10) are shown in Figure~\ref{fig:beh_ablation}. We can see that \emph{\model} consistently achieves the best performance in all cases, which verifies the significance of context behavior diversity for modeling sequential interaction patterns. We can observe that the page view and browse behaviors contribute more for purchase prediction, which shows consistent trend across two different datasets. Furthermore, we can notice that leaving behavior relation heterogeneity untapped (\ie, variant ``+buy'') will degrade the recommendation performance. Overall, the effect studies of context behavior integration reveal that the characteristics of multi-behavior data offer useful insights for modeling user sequences.\\\vspace{-0.1in}

We conduct experiments to evaluate the model performance by varying the percentages of removing user auxiliary behaviors. In particular, the percentage of 0\% indicates that we removing all behavior data under a specific type (e.g., page view or tag-as- favorite or add-to-cart); the percentage of 50\% indicates that we only keep 50\% of a specific type of user behaviors in the effect studies of context behaviors. From the new results shown in Figure 3, we can observe that incorporating more context behavior data into the multi-behavior pattern modeling is helpful in improving the performance.

\begin{figure*}
    \centering
    \begin{adjustbox}{max width=1.0\linewidth}

\begin{tikzpicture}
\begin{axis}[
    xlabel={Number of Channels $H$},
    ylabel={Hit Rate@10},
    xmin=1,xmax=17,
    ymin=0.48,ymax=0.53,
    legend columns=1,
    legend cell align=right,
    grid=both,
    every axis plot/.append style={ultra thick},
    every tick label/.append style={scale=1.3},
    label style={scale=1.8},
    legend style={
        nodes={scale=1.5, transform shape},
        legend image post style={scale=1.5},
        },
    legend style={at={(1,0)},anchor=south east},
    every axis plot post/.append style={
        every mark/.append style={scale=2}
    }
]
\addplot[color={rgb:red,133;green,76;blue,255}, mark=o, mark options={solid}]
table[x=para, y=ijcai_hr] {head-num.txt};
\legend{\large IJCAI}
\end{axis}
\end{tikzpicture}

\begin{tikzpicture}
\begin{axis}[
    xlabel={Number of Channels $H$},
    ylabel={NDCG@10},
    xmin=1,xmax=17,
    ymin=0.28,ymax=0.35,
    legend columns=1,
    legend cell align=right,
    grid=both,
    every axis plot/.append style={ultra thick},
    every tick label/.append style={scale=1.3},
    label style={scale=1.8},
    legend style={
        nodes={scale=1.5, transform shape},
        legend image post style={scale=1.5},
        },
    legend style={at={(1,0)},anchor=south east},
    every axis plot post/.append style={
        every mark/.append style={scale=2}
    }
]
\addplot[color={rgb:red,133;green,76;blue,255}, mark=o, mark options={solid}]
table[x=para, y=ijcai_ndcg] {head-num.txt};
\legend{\large IJCAI};
\end{axis}
\end{tikzpicture}

\begin{tikzpicture}
\begin{axis}[
    xlabel={Length of Sub-Users $|S_i^r|$},
    ylabel={Hit Rate@10},
    xmin=3.6,xmax=10.4,
    ymin=0.51,ymax=0.53,
    legend columns=1,
    legend cell align=right,
    grid=both,
    every axis plot/.append style={ultra thick},
    every tick label/.append style={scale=1.3},
    label style={scale=1.8},
    legend style={
        nodes={scale=1.5, transform shape},
        legend image post style={scale=1.5},
        },
    legend style={at={(1,0)},anchor=south east},
    every axis plot post/.append style={
        every mark/.append style={scale=2}
    }
]
\addplot[color={rgb:red,133;green,76;blue,255}, mark=o, mark options={solid}]
table[x=para, y=ijcai_hr] {subUsrSize_ijcai-buy.txt};
\legend{\large IJCAI}

\end{axis}
\end{tikzpicture}

\begin{tikzpicture}
\begin{axis}[
    xlabel={Length of Sub-Users $|S_i^r|$},
    ylabel={NDCG@10},
    xmin=3.6,xmax=10.4,
    ymin=0.32,ymax=0.34,
    legend columns=1,
    legend cell align=right,
    grid=both,
    every axis plot/.append style={ultra thick},
    every tick label/.append style={scale=1.3},
    label style={scale=1.8},
    legend style={
        nodes={scale=1.5, transform shape},
        legend image post style={scale=1.5},
        },
    legend style={at={(1,0)},anchor=south east},
    every axis plot post/.append style={
        every mark/.append style={scale=2}
    }
]
\addplot[color={rgb:red,133;green,76;blue,255}, mark=o, mark options={solid}]
table[x=para, y=ijcai_ndcg] {subUsrSize_ijcai-buy.txt};
\legend{\large IJCAI};
\end{axis}
\end{tikzpicture}

\begin{tikzpicture}
\begin{axis}[
    xlabel={Number of GNN Layers $L$},
    ylabel={Hit Rate@10},
    xmin=-0.2,xmax=3.2,
    ymin=0.49,ymax=0.53,
    legend columns=1,
    legend cell align=right,
    grid=both,
    every axis plot/.append style={ultra thick},
    every tick label/.append style={scale=1.3},
    label style={scale=1.8},
    legend style={
        nodes={scale=1.5, transform shape},
        legend image post style={scale=1.5},
        },
    legend style={at={(1,0)},anchor=south east},
    every axis plot post/.append style={
        every mark/.append style={scale=2}
    }
]
\addplot[color={rgb:red,133;green,76;blue,255}, mark=o, mark options={solid}]
table[x=para, y=ijcai_hr] {gnnlayer.txt};
\legend{\large IJCAI}
\end{axis}
\end{tikzpicture}

\begin{tikzpicture}
\begin{axis}[
    xlabel={Number of GNN Layers $L$},
    ylabel={NDCG@10},
    xmin=-0.2,xmax=3.2,
    ymin=0.29,ymax=0.34,
    legend columns=1,
    legend cell align=right,
    grid=both,
    every axis plot/.append style={ultra thick},
    every tick label/.append style={scale=1.3},
    label style={scale=1.8},
    legend style={
        nodes={scale=1.5, transform shape},
        legend image post style={scale=1.5},
        },
    legend style={at={(1,0)},anchor=south east},
    every axis plot post/.append style={
        every mark/.append style={scale=2}
    }
]
\addplot[color={rgb:red,133;green,76;blue,255}, mark=o, mark options={solid}]
table[x=para, y=ijcai_ndcg] {gnnlayer.txt};
\legend{\large IJCAI};
\end{axis}
\end{tikzpicture}
    \end{adjustbox}
    \begin{adjustbox}{max width=1.0\linewidth}

\begin{tikzpicture}
\begin{axis}[
    xlabel={Number of Channels $H$},
    ylabel={Hit Rate@10},
    xmin=1,xmax=17,
    ymin=0.2,ymax=0.5,
    legend columns=1,
    legend cell align=right,
    grid=both,
    every axis plot/.append style={ultra thick},
    every tick label/.append style={scale=1.3},
    label style={scale=1.8},
    legend style={
        nodes={scale=1.5, transform shape},
        legend image post style={scale=1.5},
        },
    legend style={at={(1,0)},anchor=south east},
    every axis plot post/.append style={
        every mark/.append style={scale=2}
    }
]
\addplot[color={rgb:red,133;green,76;blue,255}, mark=o, mark options={solid}]
table[x=para, y=tmall_hr] {head-num.txt};
\legend{\large Taobao}
\end{axis}
\end{tikzpicture}

\begin{tikzpicture}
\begin{axis}[
    xlabel={Number of Channels $H$},
    ylabel={NDCG@10},
    xmin=1,xmax=17,
    ymin=0.0,ymax=0.3,
    legend columns=1,
    legend cell align=right,
    grid=both,
    every axis plot/.append style={ultra thick},
    every tick label/.append style={scale=1.3},
    label style={scale=1.8},
    legend style={
        nodes={scale=1.5, transform shape},
        legend image post style={scale=1.5},
        },
    legend style={at={(1,0)},anchor=south east},
    every axis plot post/.append style={
        every mark/.append style={scale=2}
    }
]
\addplot[color={rgb:red,133;green,76;blue,255}, mark=o, mark options={solid}]
table[x=para, y=tmall_ndcg] {head-num.txt};
\legend{\large Taobao};
\end{axis}
\end{tikzpicture}

\begin{tikzpicture}
\begin{axis}[
    xlabel={Length of Sub-Users $|S_i^r|$},
    ylabel={Hit Rate@10},
    xmin=4,xmax=21,
    ymin=0.4,ymax=0.5,
    legend columns=1,
    legend cell align=right,
    grid=both,
    every axis plot/.append style={ultra thick},
    every tick label/.append style={scale=1.3},
    label style={scale=1.8},
    legend style={
        nodes={scale=1.5, transform shape},
        legend image post style={scale=1.5},
        },
    legend style={at={(1,0)},anchor=south east},
    every axis plot post/.append style={
        every mark/.append style={scale=2}
    }
]
\addplot[color={rgb:red,133;green,76;blue,255}, mark=o, mark options={solid}]
table[x=para, y=tmall_hr] {subUsrSize_tmall-buy.txt};
\legend{\large Taobao}

\end{axis}
\end{tikzpicture}

\begin{tikzpicture}
\begin{axis}[
    xlabel={Length of Sub-Users $|S_i^r|$},
    ylabel={NDCG@10},
    xmin=4,xmax=21,
    ymin=0.23,ymax=0.28,
    legend columns=1,
    legend cell align=right,
    grid=both,
    every axis plot/.append style={ultra thick},
    every tick label/.append style={scale=1.3},
    label style={scale=1.8},
    legend style={
        nodes={scale=1.5, transform shape},
        legend image post style={scale=1.5},
        },
    legend style={at={(1,0)},anchor=south east},
    every axis plot post/.append style={
        every mark/.append style={scale=2}
    }
]
\addplot[color={rgb:red,133;green,76;blue,255}, mark=o, mark options={solid}]
table[x=para, y=tmall_ndcg] {subUsrSize_tmall-buy.txt};
\legend{\large Taobao};
\end{axis}
\end{tikzpicture}

\begin{tikzpicture}
\begin{axis}[
    xlabel={Number of GNN Layers $L$},
    ylabel={Hit Rate@10},
    xmin=-0.2,xmax=3.2,
    ymin=0.12,ymax=0.5,
    legend columns=1,
    legend cell align=right,
    grid=both,
    every axis plot/.append style={ultra thick},
    every tick label/.append style={scale=1.3},
    label style={scale=1.8},
    legend style={
        nodes={scale=1.5, transform shape},
        legend image post style={scale=1.5},
        },
    legend style={at={(1,0)},anchor=south east},
    every axis plot post/.append style={
        every mark/.append style={scale=2}
    }
]
\addplot[color={rgb:red,133;green,76;blue,255}, mark=o, mark options={solid}]
table[x=para, y=tmall_hr] {gnnlayer.txt};
\legend{\large Taobao}
\end{axis}
\end{tikzpicture}

\begin{tikzpicture}
\begin{axis}[
    xlabel={Number of GNN Layers $L$},
    ylabel={NDCG@10},
    xmin=-0.2,xmax=3.2,
    ymin=0.05,ymax=0.3,
    legend columns=1,
    legend cell align=right,
    grid=both,
    every axis plot/.append style={ultra thick},
    every tick label/.append style={scale=1.3},
    label style={scale=1.8},
    legend style={
        nodes={scale=1.5, transform shape},
        legend image post style={scale=1.5},
        },
    legend style={at={(1,0)},anchor=south east},
    every axis plot post/.append style={
        every mark/.append style={scale=2}
    }
]
\addplot[color={rgb:red,133;green,76;blue,255}, mark=o, mark options={solid}]
table[x=para, y=tmall_ndcg] {gnnlayer.txt};
\legend{\large Taobao};
\end{axis}
\end{tikzpicture}
    \label{fig:seq_length}
    \end{adjustbox}
    \caption{Hyperparameter study of the \model\ framework on both Taobao and IJCAI datasets.}
    \label{fig:hyperparam}
\end{figure*}
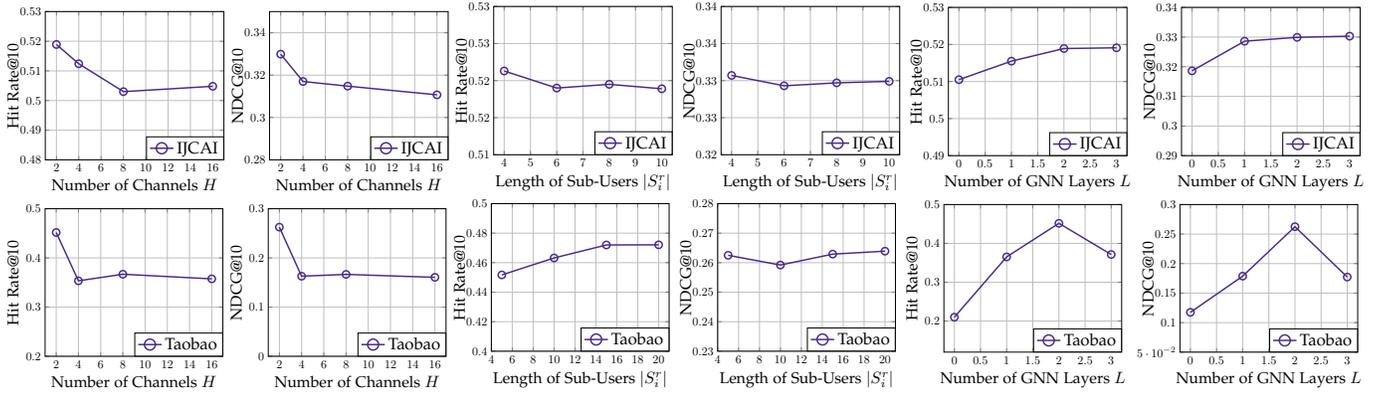

\subsection{Study on Model Hyperparameters (RQ4)}
To explore the effects of modeling dynamic multi-behavior patterns, we study how three key hyperparameters (\ie, number of projection channels $H$, the length of sub-sequence $|S_i^r|$, the depth of graph model $L$) affect the performance of our \emph{\model} framework. We show the empirical study results in Figure~\ref{fig:hyperparam} and draw the following conclusions:

 

\subsubsection{\bf Impact of Projection Channels}
In the component for encoding multi-behavior patterns, we design a multi-channel parameter learning scheme for embedding transformation. From the evaluation results shown in Figure~\ref{fig:hyperparam}, we can observe that the best recommendation performance can be achieved with the configuration of two projection channels. When we further increase the number of channels $\geq 4$, the performance becomes worse due to the overfitting problem.

\subsubsection{\bf Impact of Interaction Sub-sequence Length}
In our \emph{\model}, the interaction sub-sequence $S_i^r$ serve as the sequence unit for short-term multi-behavior pattern modeling. From the results shown in Figure~\ref{fig:hyperparam}, we observe that the influence of encoded sequence length may vary by datasets. In particular, the performance on IJCAI-Contest data is not very sensitive to the sequence length. The reason is that the short-term item-transitions could also be captured through the high-order information propagation paradigm over the constructed large-scale graph. When the sub-sequence is short, the model cannot achieve better performance, because the involved interaction behaviors are sparse. \\\vspace{-0.1in}

\subsubsection{\bf Impact of Graph Model Layers}
In our proposed \emph{\model} framework, the number of graph neural network layers is searched in the range of \{1,2,3\}. We can observe that increasing the number of embedding propagation layers to 2 (\emph{\model}-2) could substantially improve the performance over \emph{\model}-1 (with the relation learning over one-hop connections). Such performance gain can be attributed to the exploration of high-order multi-behavior patterns as well as the underlying cross-sequence relations. When we stack more propagation layers for information propagation, the performance of \emph{\model}-3 begins to deteriorate. The reason of such observation lies in the over-smoothing issue with more message passing layers, which is consistent with the findings in recent graph neural networks~\cite{li2018deeper,liu2020towards}.


\subsection{Case Study of \emph{\model} (RQ5)}
\begin{figure}[h]
    \centering
    \includegraphics[width=\columnwidth]{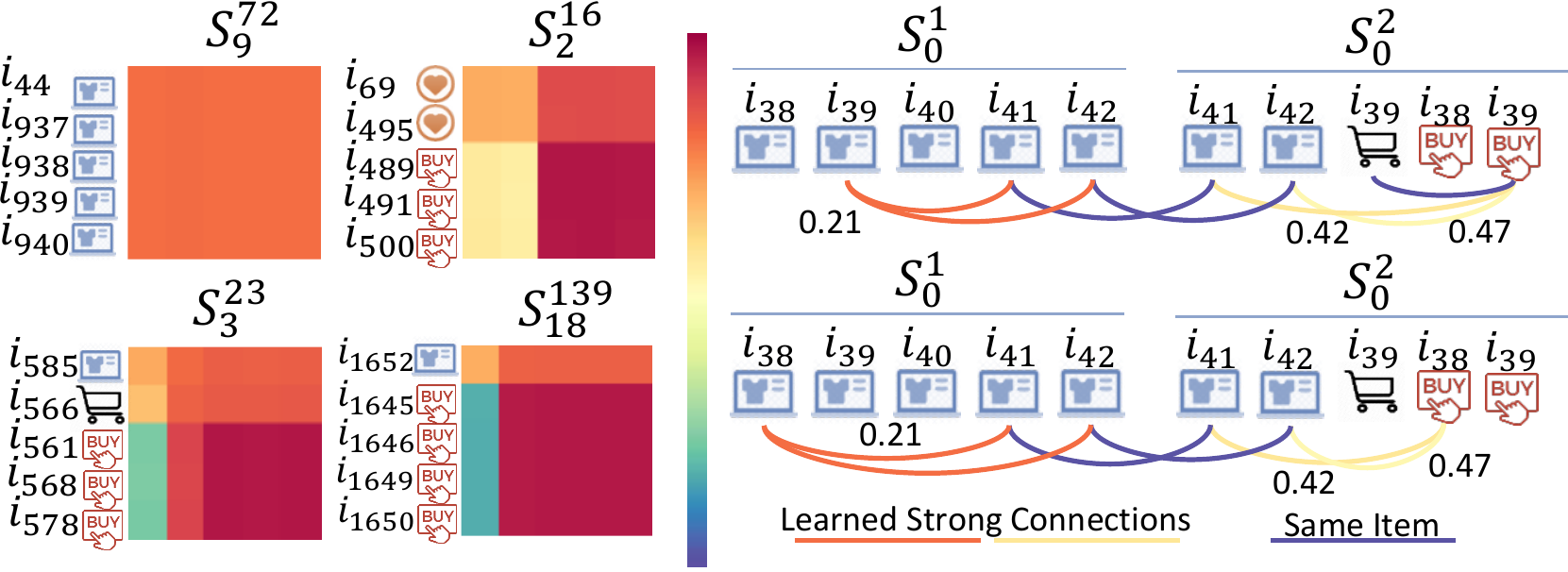}
    \caption{Model interpretation with the case studies on the learned multi-behavior relations from both short- and long-term perspectives.}
    \label{fig:case_study}
    \vspace{-0.05in}
\end{figure}

In this section, we conduct case studies on Taobao data to offer an intuitive impression of our model explainability from the perspectives of both short-term and long-term multi-behavior pattern modeling in our sequential recommender. The results are shown in Figure~\ref{fig:case_study}. The details are as follows:

\begin{itemize}[leftmargin=*]


\item We sample several users and their sub-sequences as concrete examples. For example, we can observe that the purchase behavior of user $u_3$ on item $i_{578}$ shows dependence on his/her recent activities (in the 23-th sub-sequence $S_3^{23}$) with the learned different relevance scores, such as the view behavior on item $i_{585}$, add-to-cart behavior on item $i_{566}$, and purchase behavior on item $i_{561}$. Hence, the behavior-aware item correlations learned by our \emph{\model} model can show the explicit relevance scores among different types of short-term behaviors pertinent to the user's target behavior.\\\vspace{-0.1in}

\item We present the learned item-wise dependencies encoded by our method across different user sub-sequences. From the sampled user cases, we can observe that the purchase behavior of user $u_0$ on item $i_{39}$ is correlated with his/her view activity on item $i_{41}$ and item $i_{41}$. The dependencies between view behaviors on item $i_{41}$ and item $i_{41}$ across sub-sequences $S_0^1$ and $S_0^2$ can also be captured with the explicit weights by our global relational context learning in \emph{\model}. Therefore, the dynamic long-term behavior-aware relationships between items can be preserved through our global message passing scheme.


\end{itemize}

\subsection{Model Efficiency Test}

We conduct experiments to measure the training time of our proposed TGT method and several representative baselines on both Taobao and IJCAI datasets. We report the results in Table~\ref{tab:efficiency}. Evaluations are conducted on the machine with the configuration of NVIDIA TITAN RTX GPU, Xeon W-2133 CPU. We can observe that our method \model\ can achieve comparable model efficiency when competing with state-of-the-art recommendation techniques based on graph neural networks in terms of model training time. This indicates the scalability of our \model\ in handling large-scale datasets.

\begin{table}[h]
    \centering
    \footnotesize
    \setlength{\tabcolsep}{1.2mm}
    \caption{Model training time per epoch (measured by seconds) on Taobao and IJCAI data.}
    \label{tab:efficiency}
    \begin{tabular}{lcc|lcc}
        \hline
        Model & Taobao & IJCAI & Model & Taobao & IJCAI \\
        \hline
        \hline
        MATN & 13.6 & 39.7 & GCGNN & 35.5 & 80.9\\
        HGT & 22.7 & 41.4 & MAGNN & 32.4 & 86.7 \\
        MBGCN & 23.8 & 56.3 & DeepFM & 40.5 & 105.8 \\
        \cline{4-6}
        MGNN & 34.8 & 72.0 & \emph{\model} & 32.0 & 83.1\\
        \hline
    \end{tabular}
\end{table}

\section{Conclusion and Discussion}
\label{sec:conclusion}

In this paper, we propose a novel learning framework (\model) for sequential recommendation by decomposing interactions with behavior heterogeneity. The goal of \model\ is to aggregate dynamic relation contextual signals from different types of user behaviors and generate contextualized representations for making predictions on target behaviors. Extensive experiment results demonstrate the superiority of \model\ as compared to state-of-the-art baselines, and verify the rationality of incorporating multi-behavioral patterns in performing sequential modeling of user-item interactions. Moreover, our evaluation also provides ablation study and efficiency study to further show the model effectiveness.\\\vspace{-0.12in}

In this work, we enhance the sequential recommender system with the consideration of multi-typed user-item interactions for user preference learning. There exist several promising research directions for future investigation. For example, future work could involve the investigation of external user and item features (\eg, user profile and location, item multi-modal features) to improve the recommendation performance for online user behavior modeling scenarios. In addition, how to effectively alleviate the bias factors which confounds the user preference learning for recommender systems, is also desirable to explore in future work.

\section*{Acknowledgments}
We thank the reviewers for their valuable feedback and comments. This research work is supported by the research grants from the Department of Computer Science \& Musketeers Foundation Institute of Data Science at the University of Hong Kong (HKU). The research is also supported by National Nature Science Foundation of China (62072188), Major Project of National Social Science Foundation of China (18ZDA062), Science and Technology Program of Guangdong Province (2019A050510010).

\clearpage

\bibliographystyle{abbrv}
\bibliography{refs_full}



\begin{IEEEbiography}[{\includegraphics[width=1.1in,height=1.15in,clip,keepaspectratio]{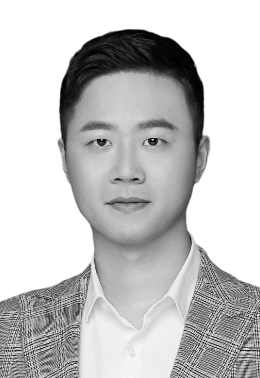}}]{Lianghao Xia}
    is currently a postdoctoral fellow in the Department of Computer Science \& Musketeers Foundation Institute of Data Science, at the University of Hong Kong. He received his B.E. and PhD degrees from South China University of Technology in 2017 and 2021, respectively. His research interests include data mining, graph neural networks and recommender systems. His research work has appeared in several major international conferences and journals such as SIGIR, AAAI, IJCAI, ICDE, CIKM, ICDM as well as ACM TOIS.\vspace{-0.1in}
\end{IEEEbiography}

\begin{IEEEbiography}[{\includegraphics[width=1in,height=1.15in,clip,keepaspectratio]{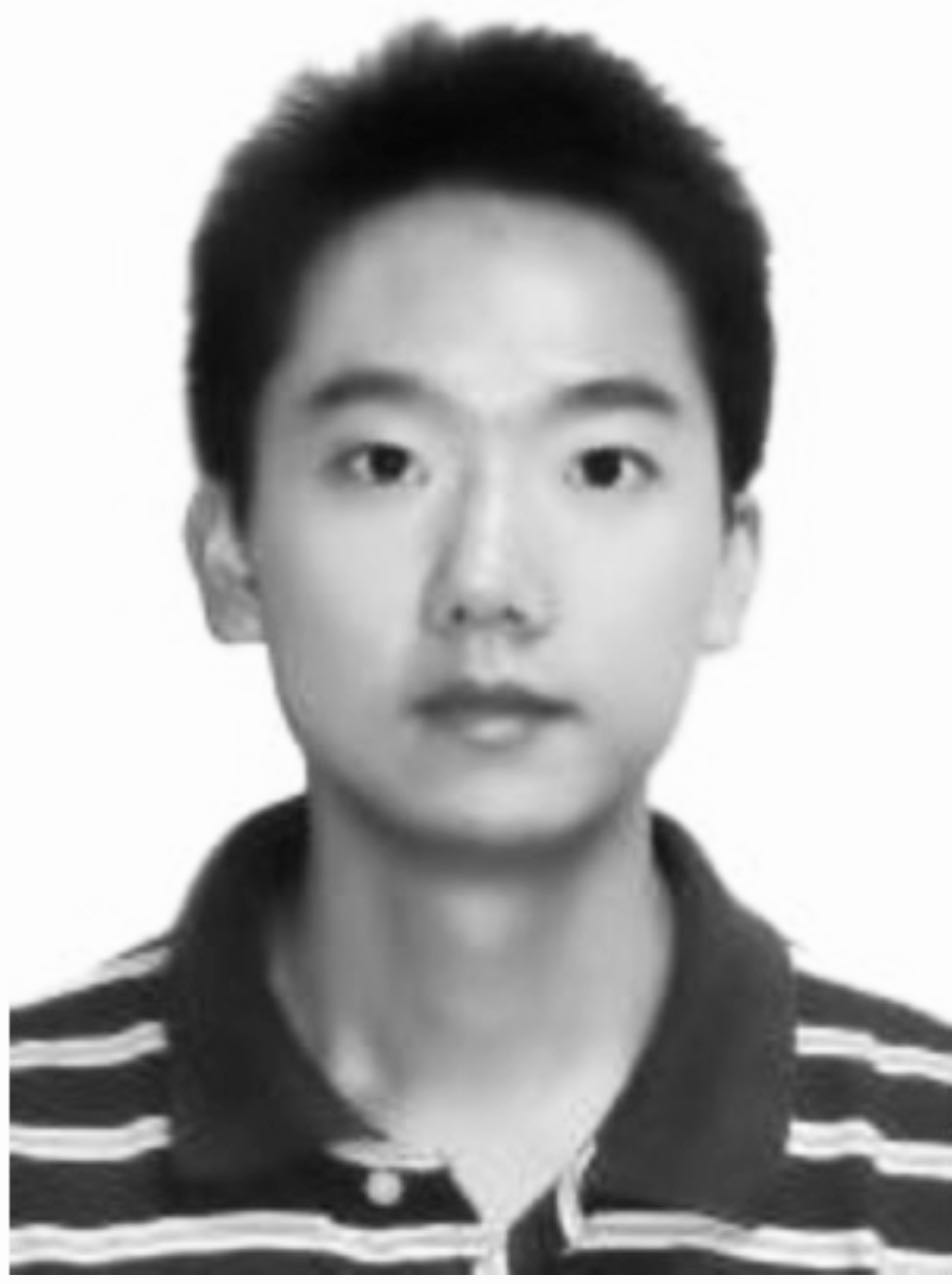}}]{Chao Huang}
	is a tenure-track assistant professor in the Department of Computer Science \& Musketeers Foundation Institute of Data Science, at the University of Hong Kong. He obtained the PhD degree from the University of Notre Dame. His research focuses on applied machine learning, graph neural networks, recommendation and spatial-temporal data mining. His work has appeared in several major international conferences and journals such as KDD, WWW, SIGIR, IJCAI, AAAI, WSDM and etc. He has served as the PC member for several top conferences including KDD, WWW, SIGIR, WSDM, AAAI, IJCAI, NIPS, ICLR and etc. Additionally, he has been recognized as the outstanding reviewer in both ACM WSDM'2020 and WSDM'2022 conference.\vspace{-0.1in}
\end{IEEEbiography}

\begin{IEEEbiography}[{\includegraphics[width=1in,height=1.1in,clip,keepaspectratio]{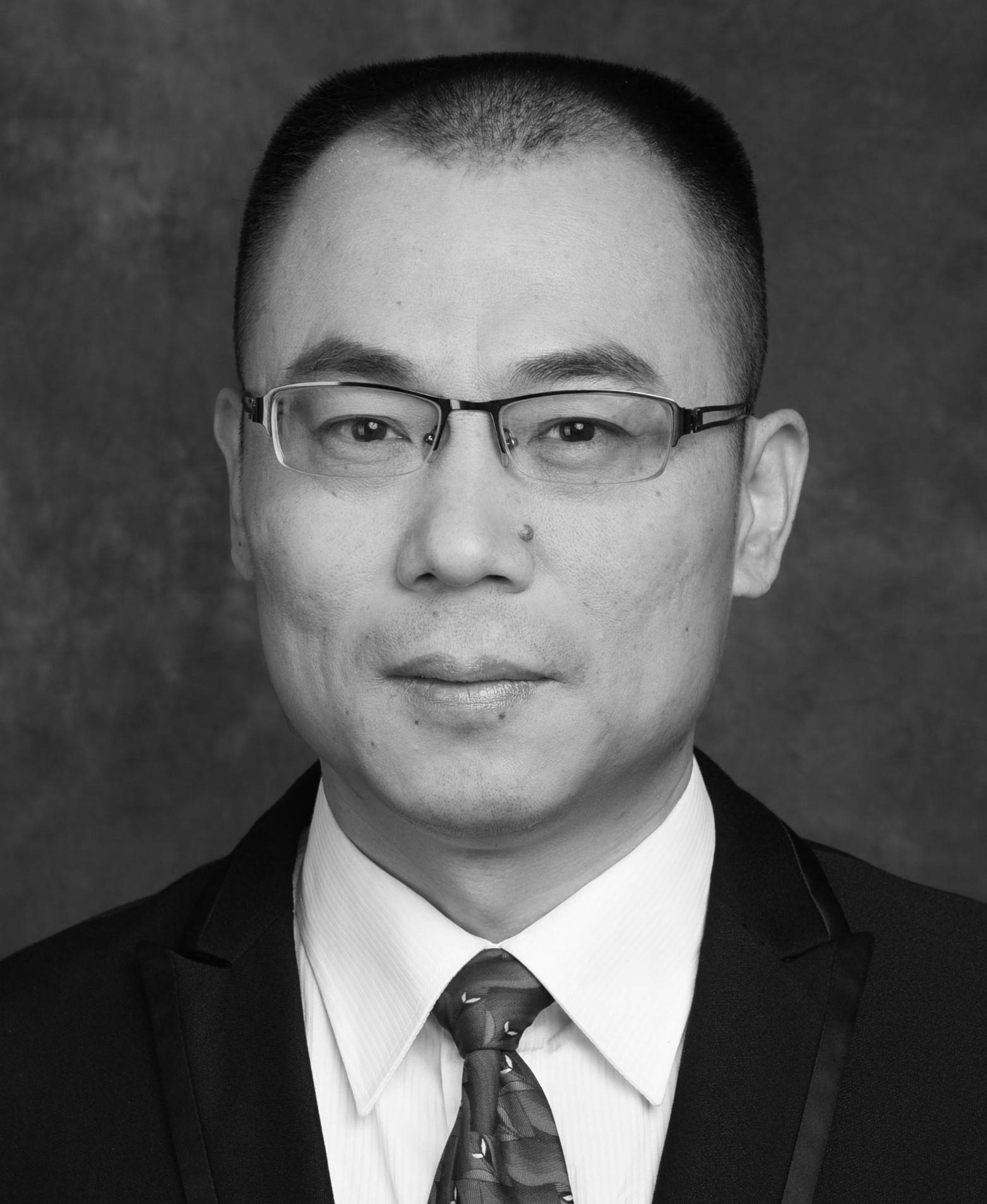}}]{Yong Xu}
is a Professor at the School of Computer Science and Engineering in South China University of Technology. His research interests include machine learning, pattern recognition and big data analysis. He has published over 80 research papers in refereed journals and conferences (\eg, SIGIR, AAAI, IJCAI, CIKM, CVPR, NIPS, ICCV, TIP, TMM and TOIS) and been serving as PC for conferences \& journals including AAAI, CVPR, ICCV, TIP and etc. Dr. Xu is a member of the IEEE Computer Society and the ACM.
\end{IEEEbiography}

\begin{IEEEbiography}[{\includegraphics[width=1.1in,height=1.25in,clip,keepaspectratio]{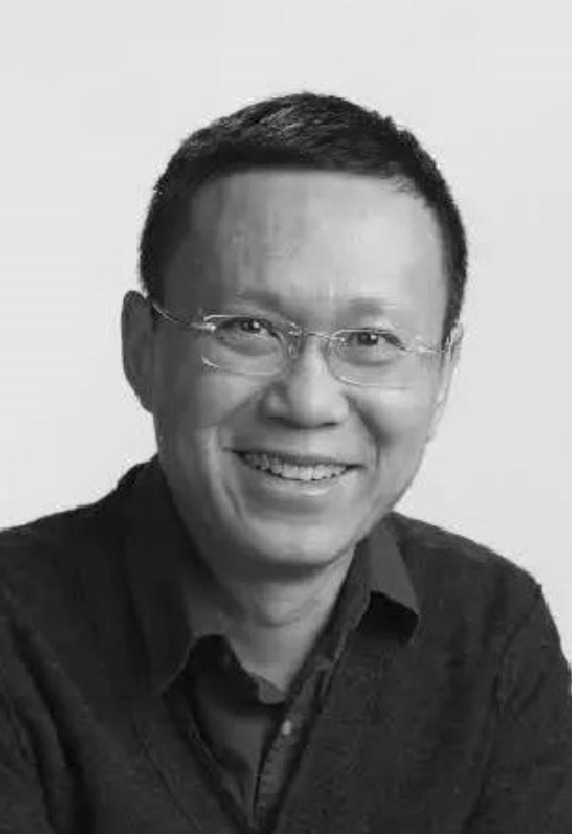}}]{Jian Pei} is currently a Canada research chair (Tier 1) in big data science and a professor with the School of Computing Science, Simon Fraser University, Canada. He is one of the most cited authors in data mining, database systems, and information retrieval. He has published prolifically and served regularly for the leading academic journals and conferences in his fields. Since 2000, he has published one textbook, two monographs and over 200 research papers in refereed journals and conferences, which have been cited by more than 107,186 in literature. He was the editor-in-chief of the IEEE Transactions of Knowledge and Data Engineering (TKDE) in 2013-2016. He is a fellow of the Association for Computing Machinery (ACM) and the Institute of Electrical and Electronics Engineers (IEEE).

\end{IEEEbiography}

\IEEEdisplaynontitleabstractindextext

%
\IEEEpeerreviewmaketitle

\ifCLASSOPTIONcaptionsoff
  \newpage
\fi



%
%
%


\end{document}